\documentclass[sigconf]{acmart}

\usepackage{xcolor}
\usepackage{tcolorbox}
\usepackage{algorithm}
\usepackage{algpseudocode}
\usepackage{tikz}
\usetikzlibrary{trees}
\usetikzlibrary{positioning}
\usepackage{fancyvrb}
\usepackage{enumitem}
\usepackage{float}
\usepackage[skip=2pt,font=footnotesize]{caption}
\usepackage{listings}
\usepackage{subfigure}
\usepackage{graphicx}
\usepackage{lipsum}

\newcommand{\tool}{\textsc{Kedavra}}
\newcommand{\sotaa}{\textsc{Arvada}}
\newcommand{\sotat}{\textsc{Treevada}}

\newcommand\blfootnote[1]{%
  \begingroup
  \renewcommand\thefootnote{}\footnote{#1}%
  \addtocounter{footnote}{-1}%
  \endgroup
}

\AtBeginDocument{%
  }

\setcopyright{acmlicensed}
\copyrightyear{2024}
\acmYear{2024}
\acmDOI{XXXXXXX.XXXXXXX}

\acmConference[ASE '24]{39th IEEE/ACM International Conference on Automated Software Engineering}{Oct 27-- Nov 1,
  2024}{Sacramento, California, United States}
\acmISBN{978-1-4503-XXXX-X/18/06}

\begin{document}

\title{Incremental Context-free Grammar Inference in Black Box Settings}

\author{Feifei Li$^\dagger$$^\ddagger$, Xiao Chen$^\S$, Xi Xiao$^\dagger$$^\ddagger$$^\ast$, Xiaoyu Sun$^\mathparagraph$, Chuan Chen$^\ddagger$, Shaohua Wang$^\|$$^\ast$, Jitao Han$^\|$} 
\affiliation{%
  \institution{$^\dagger$Tsinghua Shenzhen International Graduate School, China}
  \country{}
}
\affiliation{%
  \institution{$^\ddagger$Key Laboratory of Computing Power Network and Information Security, Ministry of Education, \\Qilu University of Technology (Shandong Academy of Sciences), China}
  \country{}
}
\affiliation{%
  \institution{$^\S$The University of Newcastle, Australia}
  \country{}
}

\affiliation{%
  \institution{$^\mathparagraph$The Australian National University, Australia}
  \country{}
}

\affiliation{%
  \institution{$^\|$Central University of Finance and Economics, China}
  \country{}
}
\renewcommand{\shortauthors}{Li et al.}

\begin{abstract}

Black-box context-free grammar inference presents a significant challenge in many practical settings due to limited access to example programs. The state-of-the-art methods, \sotaa{} and \sotat{}, employ heuristic approaches to generalize grammar rules, initiating from flat parse trees and exploring diverse generalization sequences. We have observed that these approaches suffer from low quality and readability, primarily because they process entire example strings, adding to the complexity and substantially slowing down computations. To overcome these limitations, we propose a novel method that segments example strings into smaller units and incrementally infers the grammar. Our approach, named \tool{}, has demonstrated superior grammar quality (enhanced precision and recall), faster runtime, and improved readability through empirical comparison. 
\blfootnote{*Corresponding authors: Xi Xiao and Shaohua Wang.\\\raggedright Emails: Feifei Li (lff23@mails.tsinghua.edu.cn), Xiao Chen (xiao.chen@newcastle.edu.au), Xi Xiao (xiaox@sz.tsinghua.edu.cn), Xiaoyu Sun (xiaoyu.sun1@anu.edu.au), Chuan Chen (chenchuan2019@qlu.edu.cn), Shaohua Wang (davidshwang@ieee.org), Jitao Han (hanjitao1@gmail.com).}

\end{abstract}

\maketitle

\section{Introduction}

Inferring context-free grammars (CFG)~\cite{cremers1975} from example strings~\cite{sakakibara1992,sakakibara1990,sakakibara2000} is fundamental to various software engineering tasks, including code understanding \cite{oda2015learning}, reverse engineering \cite{moonen2001generating},  protocol specification \cite{caballero2007polyglot,narayan2015survey},
detecting and refactoring code smells \cite{kim2013specification,nagy2017static}, and generation of randomized test inputs~\cite{gopinath2019building,aschermann2019nautilus,alsaeed2023finding,godefroid2008grammar,nguyen2020mofuzz,srivastava2021gramatron,wang2019superion}. Traditional methods for grammar inference often rely on grey-box or white-box approaches, where internal structures of parsers are accessible \cite{blazytko2019grimoire,lin2008deriving,hoeschele2016mining}. However, in many real-world scenarios, only black-box parsers are available, making context-free grammar inference particularly challenging. 

In a black-box setting, practitioners are typically provided with a set of example strings that adhere to an underlying grammar. Accompanying these examples is a closed-source parser, often referred to as an oracle, which has the capability to assess and verify the validity of input strings against the established grammar. The primary challenge in this environment is to infer the complete grammar based solely on the limited information provided by the example strings and the binary feedback from the oracle. This task involves constructing a robust and comprehensive representation of the grammar that can accurately predict the oracle’s validation outcomes for any given input, essentially replicating the oracle’s rule set without access to its internal logic or structure. 

Recent advancements in black-box CFG inference include \sotaa{}~\cite{kulkarni2021learning} and \sotat{}~\cite{arefin2024fast}. \sotaa{} approaches the task by inferring grammar directly from the entirety of input example strings and by exploring various generalization sequences. This method, however, introduces several significant drawbacks: notably, low accuracy, due to difficulties in detecting over-generalization within complex grammars; reduced processing speed; and diminished grammar readability, resulting from complex grammar structures. On the other hand, \sotat{}~\cite{arefin2024fast} has made improvements over \sotaa{}~\cite{kulkarni2021learning} by implementing common language concept nesting rules to pre-structure input example strings, aiming to enhance the structural predictability and efficiency of the inference process. Despite these improvements, \sotat{}~\cite{arefin2024fast} still faces the core challenges of low accuracy, slow processing speeds, and limited readability that are inherent in the methods used by its predecessor.

We propose \tool{} to address the limitations identified in previous works. Firstly, \tool{} incorporates a data decomposition step that simplifies the processing of complex data into more simple units. This enables incremental construction of the grammar, ensuring that our results are robust and less susceptible to variations in the dataset. Such decomposition not only facilitates more stable grammar inference but also enhances processing speed, particularly with complex datasets.
Secondly, \tool{} employs an incremental approach to grammar inference, systematically constructing the grammar from the simplest to the most complex structures. This method significantly improves the readability of the resulting grammar and simplifies the identification of potential over-generalizations during the inference process. Consequently, the accuracy of the grammar is substantially improved.

In summary, this paper presents several significant contributions:
\begin{itemize}
    \item We introduce \tool{}, a novel approach for CFG inference in black-box environments. The inferred grammars demonstrate robust stability and exhibit minimal sensitivity to variations in the example strings, ensuring reliable performance across diverse datasets.
    \item Through rigorous empirical evaluation, we compare the performance of \tool{} against the state-of-the-art approaches, \sotaa{} and \sotat{}. Our results indicate that \tool{} significantly outperforms these benchmarks in terms of grammar quality, readability, consistency, and computational efficiency.
    \item We proposed a more reliable sampling algorithm that can be used to evaluate other grammar inference tools.
    \item We have made the source code of \tool{} and all associated experimental results publicly available\footnote{\url{https://github.com/Sinpersrect/kedavra}} 
\end{itemize}

\section{Background}

\subsection{Context-free grammar}

Context-Free Grammar (CFG)\cite{cremers1975} is a mathematical model used to describe the grammar structure of formal languages. Formally, a context-free grammar $G$ can be represented as a quadruple $(N, \Sigma, P, S)$, where $N$ is a finite set of non-terminal symbols, typically denoted by uppercase letters; $\Sigma$ is a finite set of terminal symbols, which constitute the actual strings, or the alphabet of the language, typically denoted by lowercase letters or specific characters; $P$ is a finite set of production rules, each rule taking the form $A \rightarrow \alpha$, where $A$ is a non-terminal symbol and $\alpha$ is a string composed of terminals and/or non-terminals; and $S$ is the start symbol, a special non-terminal symbol indicating the beginning of a sentence or the start of the grammar.

Consider a simple context-free grammar defined as follows: the non-terminal symbol set $N = \{S, A\}$, the terminal symbol set $\Sigma = \{a, b\}$, and the production rules $S \rightarrow aA$, $A \rightarrow Sb$, $A \rightarrow ab$, with the start symbol $S$. Through these rules, strings such as "aab", "aaabb", "aaaabbb", etc., can be generated. The generation process begins with the start symbol $S$, then applies the rules iteratively until a string of terminal symbols is formed. For instance, starting with $S$, we can apply the rule $S \rightarrow aA$ to obtain $aA$, then apply $A \rightarrow Sb$ to get $aSb$, further apply $S \rightarrow aA$ to yield $aaAb$, and finally apply $A \rightarrow ab$ to generate $aaabb$.

The importance of context-free grammars lies in their sufficiently powerful expressiveness to represent the grammar of most programming languages; in fact, almost all programming languages are defined using context-free grammars. On the other hand, context-free grammars are simple enough that we can construct efficient parsing algorithms to verify whether a given string is generated by a specific context-free grammar. 
In real-world applications, it could be used to validate the grammar of some structured data formats, such as JSON\cite{smith2015} and XML\cite{marchal2002}. In addition, compilers or interpreters of programming languages such as C++\cite{cpp}, Java\cite{java} and JavaScript\cite{javascript} can use it to check whether the code conforms to the grammar when compiling.

\subsection{Black-box Grammar Inference}

Black-box grammar inference involves inferring the context-free grammar of a programming language by analyzing a set of example strings (\textit{e.g.}, sample programs) without accessing the internal workings of the parser. In black-box grammar inference, the parser is treated as a "black box", where only the inputs (\textit{i.e.}, example strings) and outputs (\textit{i.e.}, parsing results) are visible, while its internal state, rules, or implementation details remain hidden. In contrast, white-box grammar inference provides full access to the parser's internal logic and implementation, allowing for inspection and modification of the parser code to understand how it processes inputs and to directly extract detailed grammar rules. Gray-box grammar inference falls between the two, offering limited internal access that may allow some form of inspection or intermediate state access. 

Black-box grammar inference is particularly important in the following scenarios:
\textbf{Closed-Source Parsers}: Many programming language parsers are proprietary and closed-source, preventing users from accessing or modifying these parsers due to legal or intellectual property restrictions.
\textbf{Remote Parsers}: Some parsers are available only as remote services, such as online compilers or interpreters, where users can interact with the parser only through API calls or network interfaces, without access to its underlying implementation.
\textbf{Legacy Systems}: The source code of parsers in legacy systems may be lost or poorly documented, requiring black-box techniques to reconstruct the grammar.

\subsection{SOTA - \sotaa{} and \sotat{}}
\label{sec:SOTA}

\begin{figure}[t]
    \centering
    \includegraphics[width=0.9\linewidth]{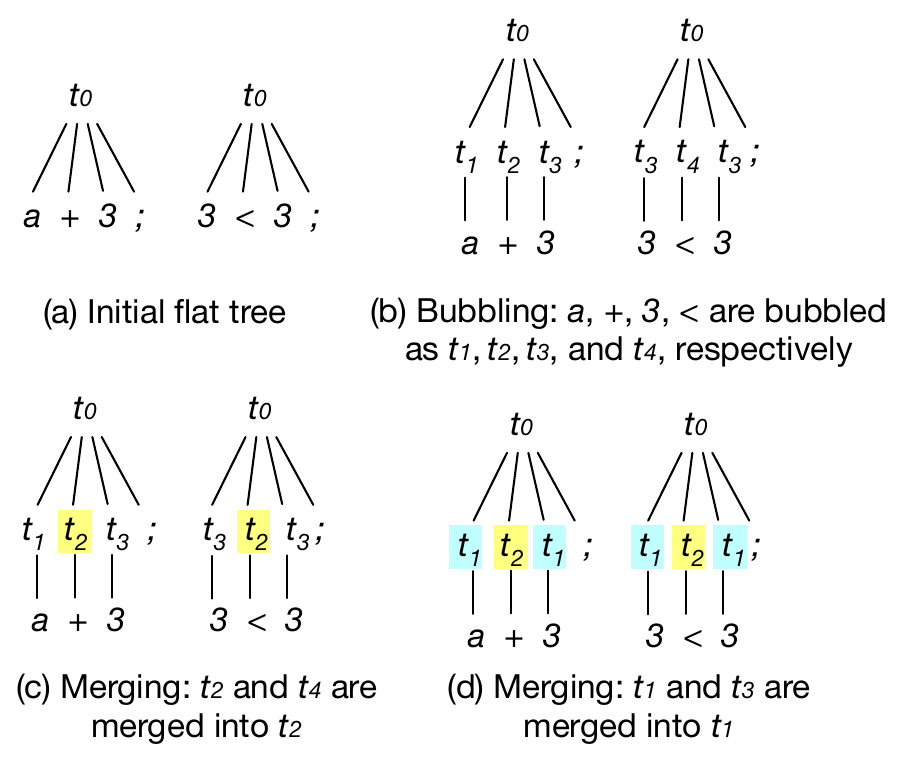}
    \caption{A simple example of \sotaa{} workflow}
    \label{fig:background}
\end{figure}

The state of the art black-box inference of context-free grammar methods include \sotaa{}~\cite{kulkarni2021learning} and \sotat~\cite{arefin2024fast}. 
To construct context-free grammar, \sotaa{} iteratively perform \textit{bubbling} and \textit{merging} operations on tree representations of the input. We demonstrate how \sotaa{} infer the context-free grammar of the input ``$a+3;$'' and ``$3<3;$'', as illustrated in Figure \ref{fig:background}. \sotaa{} first construct flat trees from the input, with a single root node $t_0$ whose children are the characters of the input strings $a+3;$ and $3<3;$ (Figure \ref{fig:background}(a)).
The bubbling operation involves taking sequences of sibling nodes in the trees (\textit{e.g.}, $a$) and adding a layer of indirection (\textit{e.g.}, $t_1$) by replacing the sequence with a new node. Then \sotaa{} decides whether to accept or reject the proposed bubble by checking whether merging the bubbles enables sound generalization of the learned language. When a bubble is merged, \sotaa{} only accepts it if it expands the language accepted by the generated grammar while preserving the Oracle validity of the strings produced by the generated grammar. Consider our example, $a$, $+$, $3$, and $<$ are bubbled as $t_1$, $t_2$, $t_3$, and $t_4$, respectively (Figure \ref{fig:background}(b)).  
If the string derivable from subtrees can be swapped and still retain a valid grammar, they can be \textit{merged}. \sotaa{} first attempts to merge bubbles that have similar context. In the above example, \sotaa{} merges $t_2$ and $t_4$ into $t_2$ because they have similar context (Figure \ref{fig:background}(c)). Then, \sotaa{} iteratively checks any remaining bubbles that can be merged. In this step, $t_1$ and $t_3$ are merged into $t_1$ (Figure \ref{fig:background}(d)). These operations are repeated until no remaining bubbled sequence enables a valid merge. Finally, the context-free grammar generated from the example is: 
\begin{Verbatim}[frame=single,commandchars=\\\{\},codes={\catcode`$=3\catcode`_=8}]
$t_{0}:t_{1} t_{2} t_{1} ``;"$
$t_{1}:``a" | ``3"$
$t_{2}: ``<" | ``+" $
\end{Verbatim}

\sotat~optimized \sotaa{} mainly by implementing predefined rules for pairing brackets.
In this example, both \sotat~and \sotaa{} yield identical results.

\section{Motivating Example}

Let us revisit the generation of the CFG of the example strings: $a+3;$ and $3<3;$, as shown in section \ref{sec:SOTA}. 
\sotat{}/\sotaa{} observes that ``+'' and ``<'' have very similar contexts, as they both precede ``3''. Therefore, \sotat{}/\sotaa{} attempts to merge ``+'' and ``<'' into $t_1$, and then merge ``a'' and ``3'' into $t_2$. 

However, this results in overgeneralization because in TinyC, ``+'' and ``<'' are not interchangeable. For example, multiple instances of ``<'' in a statement like ``a<a<a;'' are not permitted and considered illegal, whereas multiple ``+'' symbols are allowed, as seen in ``a+a+a;'', which is legal.

The method \sotat{}/\sotaa{} uses to select and merge Bubbles, specifically based on contextual similarity, is not always appropriate in certain scenarios. To address this gap, our proposed method uses an incremental construction strategy for grammar and no longer solely relies on contextual similarity when selecting and merging Bubbles. Instead, it selects Bubbles based on the generalization potential they can bring when merged. 
In addition to the issue of inaccurate generalization, \sotat{} and \sotaa{} have several other drawbacks due to inferring the grammar based on all example strings as a whole: (1) low speed, (2) low grammar readability due to complex grammar structures, (3) sensitivity of the inferred grammar to the example strings, and (4) low precision, as detecting over-generalization in complex grammar is challenging. Our proposed method addresses these drawbacks, leading to improved speed, readability, precision, robustness.

Specifically, the proposed method offers below advantages:

\textbf{More Accurate Grammar}: By evaluating the generalization potential brought by merging Bubbles, our method focuses more on the applicability and extensibility of the generated grammar. This approach enables the generated grammar to better adapt to different inputs, enhancing the model's generality and robustness, and ensuring the grammar performs well across various scenarios.

\textbf{Better Readability}: Since our Grammar is constructed incrementally, we can create a much more concise Grammar. This makes it significantly more readable compared to the previous approach.

\textbf{Faster Processing Speed}: With the incremental approach, simpler grammar that have already been processed can be skipped if they appear as part of later, more complex grammar. This approach demonstrated a significant reduction in the inference process time.

\section{Approach}

\begin{figure*}
    \centering
    \includegraphics[width=0.95\linewidth]{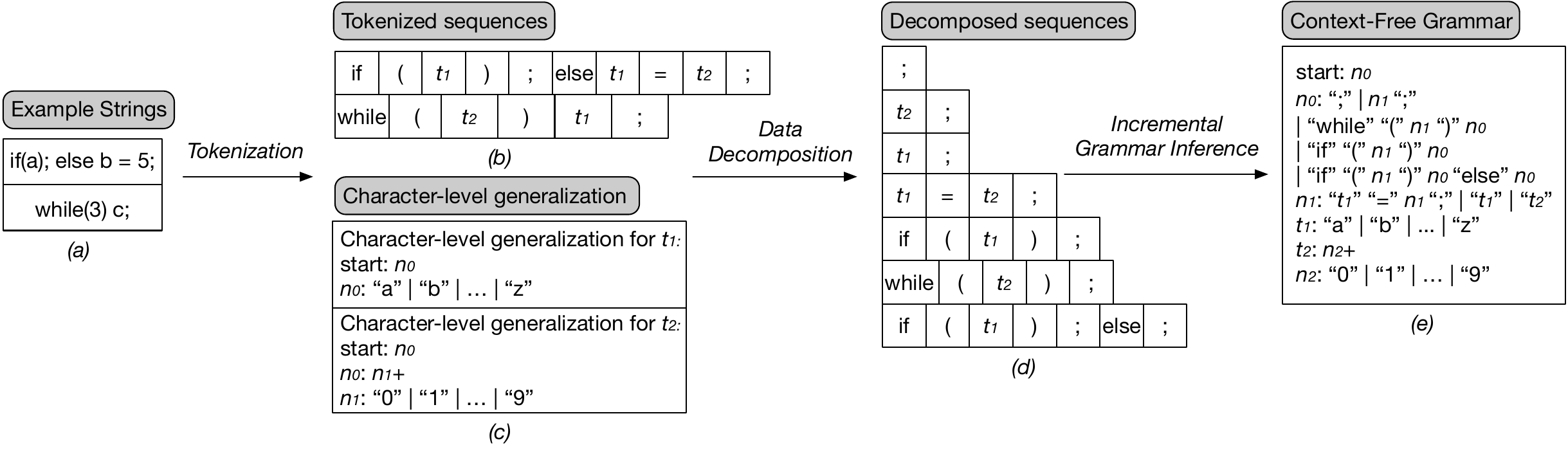}
    \caption{Workflow of \tool{}}
    \label{fig:workflow}
\end{figure*}

Figure~\ref{fig:workflow} demonstrates the workflow of the proposed \tool{}, which \textbf{consists of three main steps: tokenization, data decomposition, and incremental grammar inference.} 
\textbf{Tokenization} takes example strings as input and produce the tokenized sequences that represent syntactically significant units (\textit{e.g.}, keywords, identifiers, operators).
Tokenization accelerates subsequent data decomposition and incremental grammar inference. 
\textbf{Data decomposition} takes the tokenized sequence
as input and generates a list of decomposed sequences that break down the complex sequences into simpler ones, while collectively preserving all grammatical structures.
\textbf{Incremental grammar inference} iteratively infers the grammar from the simplest sequences to the most complex ones by decomposition, which remains significantly less complex than previous methods. This approach results in reduced complexity, improved grammar quality, and higher accuracy compared to prior works. We elaborate on the detailed steps in the following subsections.

\subsection{Tokenization}

Many languages follow fundamental tokenization rules, such as separating identifiers by non-identifier tokens.
Before performing grammar inference, \tool{} first tokenize the example strings.
\textbf{This step eliminates the need to repeatedly infer common lexing rules during grammar inference, which is computational expensive. Consequently, it significantly improves the speed of subsequent grammar inference.}

The proposed tokenization module involves the following steps.

\begin{figure}[!ht]
    \centering
    \includegraphics[width=0.7\linewidth]{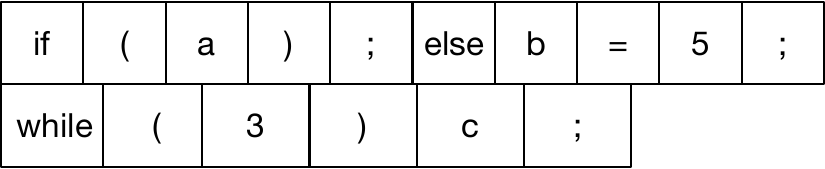}
    \caption{Results after pre-tokenization}
    \label{fig:pre-tokenization}
\end{figure}

\textbf{Pre-tokenization}: 
The proposed pre-tokenization applies common lexical rules to segment the example strings into tokens such as identifiers, strings, and numbers. \textbf{Special handling is given to whitespace.} Considering that many programming languages are not sensitive to whitespace, treating it as a token can reduce the efficiency of grammar inference (like \sotaa{} and \sotat{}). Therefore, during pre-tokenization, we also assess whether the target language is sensitive to whitespace. This involves testing whether the inputting of repeated whitespace is accepted by the oracle. If the target programming language does not treat whitespace as significant, we do not consider it as a token in the grammar inference process. An example of program before and after pre-tokenization is demonstrated in Figure \ref{fig:workflow}(a) and Figure \ref{fig:pre-tokenization}, respectively.

\textbf{Token merging}: We consider that two tokens can be merged if swapping them in all their occurrences maintains validity (to the oracle). For each pre-tokenized program, 
we attempt to merge tokens that are swappable. After merging, a token sequence and a token-value dictionary are generated. The dictionary stores the newly generated tokens and the original tokens that are merged into them. Figure \ref{fig:workflow}(b) shows an example output. 

\textbf{Character-level generalization}: The final step in tokenization involves character-level generalization of token values. We compute a character set containing all possible characters corresponding to the token\textquotesingle s character type. For instance, if the token\textquotesingle s value includes a lowercase letter, we include all lowercase letters in the set. Subsequently, we iteratively replace the original token value with each character from the set and validate the generated token value with the oracle. Furthermore, we experiment with repeating the same type of characters multiple times. For example, if the character is a lowercase letter, we attempt to repeat this letter multiple times to ascertain whether the oracle accepts the input. If the oracle accepts the generated token value, we include it in the character set. Figure \ref{fig:workflow}(c) shows an example of the character-level generalization results. 

\subsection{Data Decomposition}
\label{sec:data_decomposition}
\textbf{Previous studies, \sotaa{} and \sotat, have approached grammar inference by analyzing entire example strings. These methods often result in highly complex grammars, especially when the example strings themselves are complex.} Such complexity not only slows down the parsing process when these inferred grammars are used in parser construction but also reduces the readability of the generated grammars. The lack of readability adversely affects the user\textquotesingle s ability to understand the grammar when it is utilized for creating documentation. 

To address this shortcoming, our goal in designing the inference method is to maintain the accuracy of the grammar while making it as simple and readable as possible. To this end, \textbf{we propose breaking down complex example strings into simpler components, while collectively preserving all grammatical structures of the original program.} An example of the decomposed sequences and the corresponding input example strings are shown in Figure \ref{fig:workflow}(d) and Figure \ref{fig:workflow}(a), respectively. 

\begin{algorithm}[t]
\footnotesize
    \caption{Data decompose algorithm}\label{decompose}
    \begin{algorithmic}
    \State \textbf{Input:} tokenized sequence $D$; a oracle $O$
    \State \textbf{Output:} decomposed tokenized sequence
	\Procedure{Decompose}{$D,O$}
 
    \State {$D_{dec} \leftarrow \{\}$}
    
    \For{d in D}
    
        \For{$i \leftarrow 1\ to \ |d|$}
        
            \State {$d_{dec} \leftarrow$ keep $d[i]$ and remove other of $d$}
            
            \State {$D_{dec} \leftarrow D_{dec} \cup \{ d_{dec} \}$}
        \EndFor
        
    \EndFor
    \State \Return {$D_{dec}$}
    \EndProcedure
    \end{algorithmic}
\label{alg:data_decomposition}
\end{algorithm}

We describe the pseudocode of this step in Algorithm \ref{alg:data_decomposition}. In decomposing the data, we follow this method: First, we number all tokens in the sequence - generated from the previous step - from 1 to $n$. We then preserve the first token and attempt to eliminate as many of the remaining tokens as possible. We assess the reduced token sequence with the oracle to determine its validity; if deemed valid, we continue with the reduction; otherwise, we revert to the original sequence. This process is repeated iteratively until no additional tokens can be removed. We apply this method sequentially for each token from 1 to $n$. The result of this procedure is $n$ sequences, The $i$th sequence represents the smallest set of valid token sequences that must contain the $i$th token. As each token is preserved at least once, this approach ensures that no essential grammar elements are lost while simplifying the sequence.

Figure \ref{fig:workflow}(b) and Figure \ref{fig:workflow}(d) presents an example of the input and output from this step. The comparison clearly shows that the decomposed sequence is simpler than the original input while retaining all grammatical elements from the original input.

\subsection{Incremental Grammar Inference}
In this section, we detail our methodology for incrementally inferring grammar based on decomposed token sequences. \textbf{First, we arrange these sequences by length - from the shortest to the longest - based on the number of tokens.} Our approach incrementally incorporates the least complex sequences into the inference process.

\begin{algorithm}[t]
\footnotesize
\caption{Incremental Grammar Inference Algorithm}\label{alg:incremental_inference}
\begin{algorithmic}[1]
\State \textbf{Input:} tokenized sequences $D$;a oracle $O$
    \State \textbf{Output:} grammar
\Procedure{InferGrammar}{D,O}
\State{$G \gets \Call{createEmptyGrammar}{ }$}

\While{$D\neq \emptyset$}
    \State $T \gets$ \Call{PopShortest}{$D$}\Comment{\textcolor{gray!60}{pop the shortest token sequence in $D$}}
    \If{$T\in G$}
        \State \textbf{continue}
    \EndIf
    \State $G \gets \Call{AddProd}{G,T}$
    \While{$G$ contains production need generalize}
        \State $P \gets$ \Call{ChooseProd}{$G$}\Comment{\textcolor{gray!60}{Choose a production that need generalize.}}
        \State $B \gets$ \Call{Bubbling}{G,P,O}
        \State $B_{opt} \gets \arg\max_{i\in B}\Call{ScoringBubbleSet}{G,i,O}$
        \If{$\Call{ScoringBubbleSet}{G,B_{opt},O}=0$}\Comment{\textcolor{gray!60}{If the bubble set's \hspace*{0pt}\hfill score is 0,we skip.}}
            \State \textbf{continue}
        \EndIf
        \State $B_{filtered} \gets$ \Call{ElimOverGen}{$G,B_{opt},O$}
        \State $G \gets \Call{MergeBubbles}{G,B_{filtered}}$)
    \EndWhile
\EndWhile
\State $G \gets $ \Call{GeneralizeREP}{$G,O$}

\State\Return{G}
\EndProcedure
\end{algorithmic}
\end{algorithm}

Algorithm \ref{alg:incremental_inference} outlines our incremental inference approach. \textbf{We begin by inferring the grammar from the simplest token sequence generated in the previous step, and continue by iterating over all token sequences derived from the example strings.} The following subsections elaborate on the methodologies implemented in each iteration.

\begin{algorithm}[t]
\footnotesize
    \caption{Bubbling algorithm}\label{bubbling}
    \begin{algorithmic}
    \State \textbf{Input:} grammar G; a production P; a oracle $O$
    \State \textbf{Output:} $bubbleSets$
	\Procedure{Bubbling}{$G,P,O$}
 
    \State{$bubbleSets \leftarrow \{\}$}
    
    \For{$bubble\in \Call{SubSeq}{P}$}~\Comment{\textcolor{gray!60}{Build bubbles by iterate all \hspace*{0pt}\hfill sub sequence of $P$}}
        
        \State{$bubbleSet \leftarrow \{bubble\}$}
        
        \For{$P' \in \Call{AllProds}{G}$}
            \For{$bubble'\in \Call{SubSeq}{P'}$}
                \If{$\Call{CheckSwap}{bubble,bubble',O}$}~\Comment{\textcolor{gray!60}{Check if two bubbles \hspace*{0pt}\hfill can be swapped.}}
                    \State {$bubbleSet \leftarrow bubbleSet\cup \{ bubble' \} $}
                \EndIf
            \EndFor
        \EndFor
        
        \State $bubbleSets\leftarrow bubbleSets \cup \{ bubbleSet\}$
        
    \EndFor
    \State\Return{bubbleSets}
    \EndProcedure
    
    \end{algorithmic}
\end{algorithm}

\subsubsection{Bubbling.}

\textbf{The purpose of bubbling is to identify parts of the grammar that can be generalized.}
Algorithm \ref{bubbling} outlines how the bubbling step works.
For each
production rule $P$ that derived from the input token sequence, 
we group all its subsequences as bubbles $B_P$. Then we also group all the subsequences of the grammar's production rules as bubbles $B_{all}$ .
Here,  we employ the \sotat{}'s approach to handle parentheses. We discard bubbles with unmatched brackets, such as those containing a left bracket without a corresponding right bracket.

Then we iterate through $B_P$
and assess whether each bubble can be swapped with bubble in $B_{all}$. 
We define two bubbles as \textit{swappable} if the grammar, after the swap, can still be validated by the target oracle. 
For each bubble $B_i$ in $B_P$,we create a bubble set $\{B_i,B_j,B_{j+1}...,B_m\}$, where $B_j,B_{j+1}...,B_m$ is all bubbles in $B_{all}$ that can be swapped with $B_i$.

The generated bubble sets are then fed into the next step.

\begin{algorithm}[b]
\footnotesize
\caption{Calculate bubble set score}
\label{alg:estimate-bubble-set-score}
\begin{algorithmic}[1]
\State \textbf{Input:} grammar $G$;a bubble set $S$;a oracle $O$
    \State \textbf{Output:} bubble set score
\Procedure{ScoringBubbleSet}{G,S,O}
\State $B_{first} \gets \Call{ChooseFirstBubble}{S}$
\State $S_{remain} \gets S \setminus \{B_{first}\}$
\State $samples \gets \{\}$
\For{$B \in S_{remain}$}\Comment{\textcolor{gray!60}{Swap first bubble with each of others, and record the \hspace*{0pt}\hfill samples}}
    \State $samples \gets samples \cup \Call{SwapBubble}{B_{first}, B}$\Comment{\textcolor{gray!60}{We swap the two \hspace*{0pt}\hfill bubble, then we get two samples}}
\EndFor
\State $samples_{gen} \gets \{i \in samples \mid \Call{CheckGen}{i,G, O}\}$\Comment{\textcolor{gray!60}{Check how many \hspace*{0pt}\hfill of the swapped samples can generalize.}}
\State\Return{$|samples_{gen}|$}
\EndProcedure
\Procedure{CheckGen}{Sample,G,O}\Comment{\textcolor{gray!60}{We consider a sample that not in grammar \hspace*{0pt}\hfill but can accepted by oracle as generalized.}}
\State\Return{$Sample\not\in G\land \Call{Check}{Sample,O}$}
\EndProcedure
\end{algorithmic}
\end{algorithm}

\subsubsection{Choose the Most Generalizable Bubble Set} 
\label{sec:choose_optimal_bubble}
Algorithm \ref{alg:estimate-bubble-set-score} outlines how to calculate the bubble set's score for choosing the most generalizable bubble set.

\textbf{We choose the bubble set that has the highest generalization score, indicating the potential to produce the most generalizations.} Consider bubble set $\{B_i,B_j,B_{j+1}...,B_m\}$, We first swap $B_i$ and $B_j$, and get two sample string from respectively position. then swap $B_i$ and $B_{j+1}$ until $B_i$ and $B_{m}$ are swapped. The number of samples that result in a new generalization can serve as the generalization score. This score reflects the extent of generalization achievable by the bubble set. 

A new generalization is considered to be created if we swap two bubbles so that the resulting grammar contains a string accepted by the oracle but not included in the grammar before the merge.

We then choose the bubble set with the highest score for further processing in the next step. 

\subsubsection{Eliminating Over-generalization} 
\label{sec:filter_bubble_set}
The bubble set \{$B_i$, $B_j$, $B_{j+1}$ ... , $B_m$ \} selected in the previous step only examines whether $B_i$
are swappable with other bubbles (\textit{i.e.},$B_j, B_{j+1}..., B_m$). However, we did not check if the bubbles generated from the existing grammar (\textit{i.e.}, $B_{j+1}...,B_m$) are swappable. Merging these unswappable bubbles may introduce over-generalization, defined as the grammar containing strings that cannot be accepted by the oracle. 
In this step, we aim to select a subset of bubbles that maximize generalizability without causing over-generalization.
Algorithm \ref{alg:eliminate-over-gen} outlines our method for this step.

\begin{algorithm}[t]
\footnotesize
\caption{Eliminating Over-generalization}
\label{alg:eliminate-over-gen}
\begin{algorithmic}[1]
\State \textbf{Input:} grammar $G$;a bubble set $S$;a oracle $O$
    \State \textbf{Output:} bubble set without over-generalization
\Procedure{ElimOverGen}{$G,S,O$}
\State{$G' \gets \Call{MergeBubble}{G,S}$}
\State{$Samples\gets \{i\in\Call{Sample}{G'}\mid \Call{CheckGen}{i,G,O}\}$} \Comment{\textcolor{gray!60}{We create samples of $G'$ that indicate the full generalization, it can be used to estimate the \hspace*{0pt}\hfill generalization score of $S$'s subset}}
\State $S_{min} \gets S$
\ForAll{$b \in S$}
    \If{\Call{GenScore}{$S_{min} \setminus \{b\},G,Samples$} = 1}\Comment{\textcolor{gray!60}{If we remove a bubble \hspace*{0pt}\hfill does not reduce the generalization score, we remove it}}
        \State $S_{min} \gets S_{min} \setminus \{b\}$
    \EndIf
\EndFor

\State $subsets \gets \{S_{sub} \subseteq S_{min} \mid \Call{CheckOverGen}{S_{sub}}\}$\Comment{\textcolor{gray!60}{We choose all \hspace*{0pt}\hfill subset of $S_{min}$ that does not cause over-generalization}}
\State\Return{$\arg\max_{i\in subsets}\Call{GenScore}{i,G,Samples}$}
\EndProcedure
\Procedure{GenScore}{$B,G,Samples$}
\State{$G'\gets \Call{MergeBubbles}{G,B}$}
\State\Return{$\mid\{i\in Samples\mid i\in G'\}\mid/\mid Samples\mid$} \Comment{\textcolor{gray!60}{Check how many \hspace*{0pt}\hfill samples are in the grammar after merging Bubbles.}}
\EndProcedure
\end{algorithmic}
\end{algorithm}

First, we try to find a minimal subset with full generalization: we remove all bubbles in the subsets that do not affect the generalization score. The generalization score is calculated as follows: First we collect all the generalized samples from the grammar that merging the bubbles in section \ref{sec:choose_optimal_bubble}. Then we evaluate how many of these samples are in the grammar after merging the bubble set (the bubble set we need to estimate generalization score). This rate is the generalization score.
This process yields minimal complete generalization subsets. \textbf{Based on this subset, we seek to find a subset that 1) does not over-generalize, and 2) has the highest generalization score.}

To achieve this, we enumerate all subsets of the current bubble set, then select all subsets that do not cause over-generalization. From these, we choose the subset with the highest generalization score. 

We would like to note that \tool{} exhibits non-determinism, primarily during the process for eliminating over-generalization. In this phase, we utilize a sampling method to select grammar samples for estimating the generalization score, which introduces non-determinism into our approach.

\subsubsection{Merging and Grammar Simplification}

Merging involves uniting all bubbles into a single non-terminal. We observed that the merged grammar may introduce redundancy. To mitigate this redundancy and accelerate the inference process, we further simplify the generated grammar. For example, if a non-terminal only appears in one production rule and this production rule does not contain any other non-terminal or terminal, we merge this non-terminal with the production rules it produces and so on. It is important to note that these simplify rules were observed during the experiments and may not encompass all possible redundancies. However, new rules can be added if additional cases are identified. Notably, simplifying the grammar speeds up the inference process without impacting accuracy.

\subsubsection{Generalize Rep and Expansion of Terminals}
After iterating through all sequences generated from the example strings (as described in Section \ref{sec:data_decomposition}), we identify the repeatable bubbles within the grammar. This is achieved by sample generated with repeated bubbles against the oracle. If the sample generated by the repeated bubbles is accepted by the oracle, the bubble is marked as repeatable.

At this point, we obtain a context-free grammar inferred from the example strings.
The final step involves expanding each terminal in the grammar to a larger character class, based on the character-level generalization table produced in Section \ref{sec:data_decomposition}. This ensures that the grammar can accept tokens of the same type that were not present in the example strings
but are still valid. For instance, expanding  $t_1 \rightarrow a \ | \ b \ | \ c$  to include all lowercase letters.

\section{Evaluation}
We designed a novel algorithm with the aim of inferring as many high-quality grammars as possible. To evaluate whether this goal has been fulfilled, we applied \tool{} to 8 micro benchmarks and 3 macro benchmarks, comparing it to the most cutting-edge tools. We then provide an in-depth analysis of the comparison results to answer the following five research questions.

\textbf{RQ1} Grammar quality: Does \tool{} infer better grammars than \sotat{} and \sotaa{} on the same dataset?

\textbf{RQ2} Runtime: Is the runtime of \tool{} lower than \sotat{} and \sotaa{} on the same dataset?

\textbf{RQ3} Readability: How compact are the inferred grammars?

\textbf{RQ4} The impact of different sampling methods on grammar accuracy.

\subsection{Experimental Setup}
\begin{table}
\caption{Example strings S with their avg char size; \# = nr. programs;}\label{table:dataset}
\begin{tabular}{c|cc|cc|cc|cc}
  \hline
  & \multicolumn{2}{c|}{R0} & \multicolumn{2}{c|}{R1}  & \multicolumn{2}{c}{R2} & \multicolumn{2}{c}{R5} 
   \\
    & \# & avg & \# & avg & \# & avg & \# & avg  \\\hline
arith&17&2.3&\multicolumn{6}{c}{n/a}\\
fol&36&14.5&\multicolumn{6}{c}{n/a}\\
math&62&5.5&\multicolumn{6}{c}{n/a}\\\hline
json&71&3.9&30&11.7&30&8.6&\multicolumn{2}{c}{n/a}\\
lisp&26&2.5&30&79.2&30&24.8&\multicolumn{2}{c}{n/a}\\
turtle&33&7.7&35&41.1&35&25.6&\multicolumn{2}{c}{n/a}\\
while&10&15.5&30&171.4&30&217&\multicolumn{2}{c}{n/a}\\
xml&40&11.6&20&27.8&20&27.5&\multicolumn{2}{c}{n/a}\\\hline
curl&25&20.7&25&22.1&25&22.0&\multicolumn{2}{c}{n/a}\\
tinyc&25&80.5&25&96.2&25&86.4&10&514\\\hline
minic&10&107&\multicolumn{6}{c}{n/a}\\\hline
\end{tabular}
\end{table}

To investigate the effectiveness and efficiency of \tool{} in grammar inference, we applied our tool on
a slightly modified benchmark dataset\footnote{
https://github.com/Sinpersrect/kedavra}, which includes a total of 11 grammars, each of which contains 1k test programs. 
To this end, we include all the original 
datasets from \sotaa{}\footnote{https://github.com/neil-kulkarni/arvada} and \sotat{}\footnote{https://github.com/rifatarefin/treevada}.
We removed Nodejs from the original dataset because the Nodejs dataset's oracle was incorrectly implemented, leading to erroneous outputs. For instance, the oracle would consider "()=\}" as syntactically correct. In reality, this input is incorrect, but the oracle used Nodejs's  \emph{--check} feature and checked the output for `SyntaxError'. However, when the input contains multiple errors, other errors can obscure the \textquotesingle SyntaxError \textquotesingle. This results in the oracle incorrectly considering an erroneous input as syntactically correct. For example, "()=\}" would trigger a \textquotesingle ReferenceError \textquotesingle instead of a \textquotesingle SyntaxError\textquotesingle , causing the oracle to wrongly deem it as syntactically correct.

In addition, to compensate for the reduced dataset resulting from the removal of the Node.js dataset and to further evaluate \tool{}'s capability with sophisticated grammar features, we provided a larger and more complex C program called "minic" which includes function calls, variable declarations, and function declarations. Additionally, we utilized \tool{}'s sampling algorithm to generate test and training data for this program. At the same time, we also leveraged \tool{}'s sample method to generate test and training data for this program. Hence, the overall dataset for the experiment contains 25 train datasets correspond to 11 grammar. Table~\ref{table:dataset} shows the size and average char size of each train dataset.
Our experiment runs on a Linux server with Intel(R) Xeon(R) Silver 4210R CPU @ 2.40GHz and 128GB RAM.

To facilitate easy comparison, we conducted the experiments using \sotat{}'s Docker image and reused its blackbox parser. It's worth noting that when assessing the precision of the generated grammar, we not only attempted to replicate the previous sampling strategy as closely as possible but also constructed a new sampling strategy: 
We limit the usage of each production rule to
no more than 10 times. 
This approach addresses the issue of the previous sampling method's inability to detect deeper grammar errors.

\subsection{RQ1: Grammar Quality}
Our first research question concerns the quality of grammar generated by \tool{} and how it compares to existing tools. 
In this work, like many other grammar inference approaches, we prioritize precision, recall and F1 score to evaluate the grammar quality. Specifically, these three metrics are calculated as follows:  
\begin{itemize} [leftmargin=*]
    \item \textbf{Precision}: We sampled 1,000 cases from the inferred grammar and then assessed how many of these were accepted by the oracle to calculate the precision,
    \item \textbf{Recall}: Based on a set of 1,000 "golden" test programs, we calculate the recall by determining how many of the inferred grammar rules appear in these programs, 
    \item \textbf{F1 score}: F1 score is the harmonic mean of precision and recall.
\end{itemize}

\textbf{Result.} Our experimental results (on R0)
are presented in table~\ref{tab:Grammar_Quality_Comparison}
, which shows the performance (\textit{i.e.}, precision, recall, and F1 score) of \tool{} compared to two state-of-the-art (SOTA) tools targeting the same problem of grammar inference. Here, we chose \sotaa{}~\cite{kulkarni2021learning} and \sotat{}~\cite{arefin2024fast} as baseline because they are recognized as the most cutting-edge tools~\cite{arefin2024fast} and have been made publicly available in the community.

\begin{table}
\scriptsize
\caption{Rerun \sotaa{} (left), rerun \sotat{} (middle) and \tool{}(right) on the same example strings; \sotaa{}/\tool{} values are average over 10 runs; f1 = F1 score; t = runtime; ± = standard deviation; bold = \tool{} better than or equal to \sotaa{}\& \sotat{}}\label{tab:Grammar_Quality_Comparison}
\begin{tabular}{l|ccc|ccc|ccc}
  \hline
  \multicolumn{1}{c|}{}& \multicolumn{3}{c|}{\sotaa{}}& \multicolumn{3}{c|}{\sotat{}}  & \multicolumn{3}{c}{\tool{}} \\
  &  p & r & f1 &  p & r & f1 &  p & r & f1\\\hline
arith&1.0±.0&1.0±.0&1.0±.0&1&1&1&\textbf{1.0±.0}&\textbf{1.0±.0}&\textbf{1.0±.0}\\
math&.97±.03&.75±.2&.83±.13&1&.47&.64&.94±.03&\textbf{.92±.1}&\textbf{.93±.07}\\
fol&.99±.03&.73±.31&.8±.23&1&.53&.69&.99±.04&.55±.18&.69±.14\\\hline
json&.91±.09&.97±.09&.93±.06&.98&.94&.96&\textbf{1.0±.0}&.97±.02&\textbf{.99±.01}\\
lisp&.97±.06&.38±.26&.5±.19&.7&.73&.71&\textbf{1.0±.0}&\textbf{1.0±.0}&\textbf{1.0±.0}\\
turtle&1.0±.0&1.0±.0&1.0±.0&1&1&1&\textbf{1.0±.0}&\textbf{1.0±.0}&\textbf{1.0±.0}\\
while&1.0±.0&.75±.23&.84±.15&1&.24&.38&\textbf{1.0±.0}&\textbf{1.0±.0}&\textbf{1.0±.0}\\
xml&1.0±.0&.96±.1&.98±.06&1&1&1&\textbf{1.0±.0}&\textbf{1.0±.0}&\textbf{1.0±.0}\\\hline
curl&.58±.08&.92±.02&.71±.06&.8&.71&.75&\textbf{1.0±.0}&.11±.12&.18±.19\\
tinyc&.55±.24&.76±.36&.6±.29&.74&.9&.81&\textbf{1.0±.0}&\textbf{1.0±.0}&\textbf{1.0±.0}\\
minic&.65±.32&.29±.17&.32±.11&.63&.62&.63&\textbf{1.0±.0}&.48±.0&\textbf{.65±.0}\\
\hline
\end{tabular}
\end{table}

Overall, as highlighed in table~\ref{tab:Grammar_Quality_Comparison}, among the 
11 grammar datasets, the F1 score of \tool{} (\textit{i.e.}, \emph{arith ,math ,json ,lisp ,turtle ,while ,xml ,tinyc and minic}, respectively on average) is higher than or equals to that of \sotaa{} and \sotat{}, further indicating that our tool performs better on most of the datasets. For example, \tool{} achieves high scores on \emph{arith, lisp, turtle, while, xml , and tinyc} with a precision, recall, and F1 score of 1.0±0.0. Even for the largest program input dataset, Minic, \tool{} still maintains good performance, with a high precision of 1.0±0.0 and an F1 score of 0.65±0.0. This high score experimentally shows the effectiveness of \tool{} in generating grammar with high quality. The outliers are fol and curl, where \tool{}'s f1 score (\textit{i.e.}, 0.69 and 0.18) is below that of \sotaa{} (\textit{i.e.}, 0.8 and 0.71) and \sotat{} (\textit{i.e.}, 0.69 and 0.75). This is because in \emph{fol}, we did not merge the optimal bubbles when inferring the grammar, which caused us to lose generalization in the next step, resulting in a lower F1 score compared to \sotaa{} and \sotat{}. In \emph{curl}, our tokenization algorithm does not fit curl's grammar well, as it is not similar to a programming language. Consequently, the poor tokenization result leads to low recall in \emph{curl}. Further discussion of the performance on \emph{curl} can be found in Section \ref{sec:discussion}. On the other hand, \tool{}'s recall in \emph{minic} is lower than in \sotat{}, which can be mainly explained by  the limitations of grammar loss, where the grammar of function calls in \emph{minic} might be lost in data decomposition because we treat a function call as a combination of an identifier and an expression. During the process of data decomposition, only one of them will be retained. Overall, despite of these imprecise results, our approach is effective in generating better grammar compared to SOTA tools.

\begin{figure}
    \centering
    \includegraphics[width=1\linewidth]{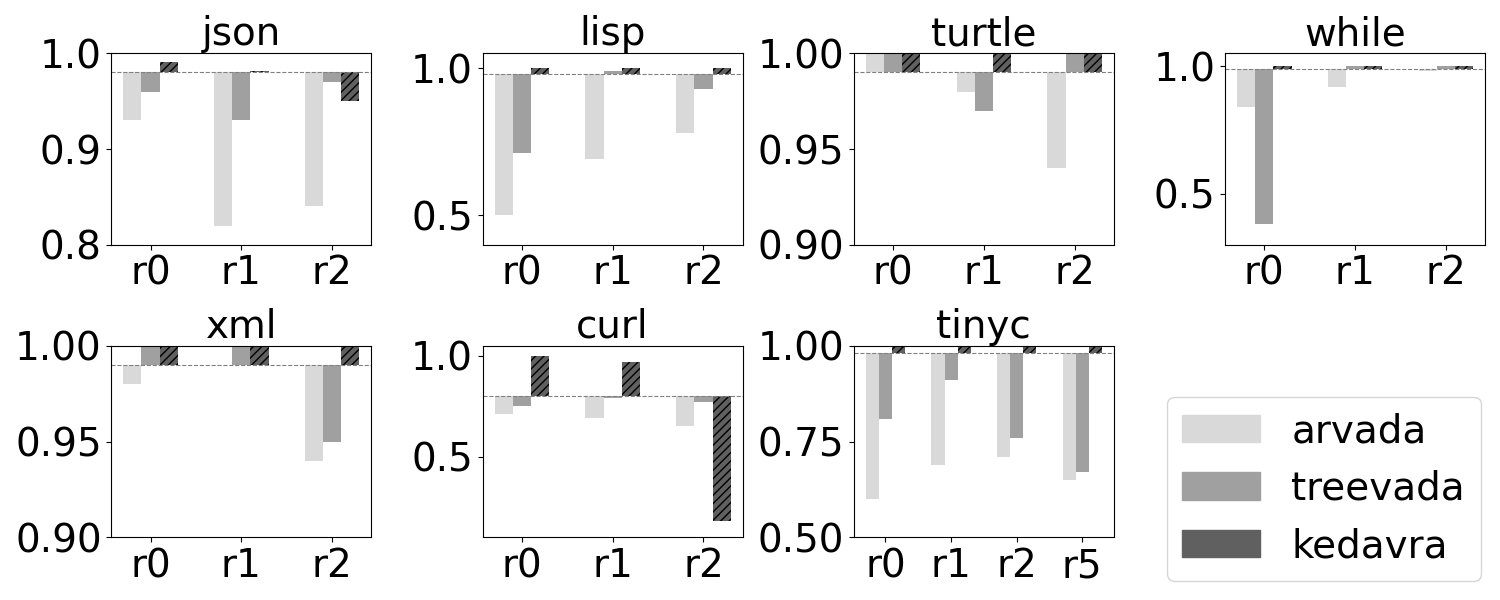}
    \caption{avg F1 score of 10 runs of \sotaa{}, \sotat{} and \tool{} on each dataset (R0, R1, R2, R5). Note that the horizontal bars in each of the sub-figures are manually added as a reference to better visualize the fluctuations in the F1 scores of the inference algorithm across different datasets.}
    \label{fig:stability}
\end{figure}

In addition to evaluating the grammar quality, it remains unknown whether \tool{} can consistently maintain consistent performance across diverse datasets. To this end, we further analyse the F1 scores on different datasets (\textit{i.e.}, R0, R1, R2, R5) with the same grammar, as presented in Figure~\ref{fig:stability}. This figure displays the F1 score of three tools across seven grammar datasets, using the distance of the bars from the horizontal axis to represent the deviation of the F1 scores. Note that the horizontal bars in each of the sub-figures are manually added as a reference to better visualize the fluctuations in the F1 scores of the inference algorithm across different datasets. Intuitively, we find that, our approach shows relatively small variations (\textit{i.e.}, \tool{} has the shortest distance from the horizontal axis) in the learned results, whereas previous methods are heavily influenced by dataset characteristics. Across various grammars such as Lisp, Turtle, TinyC, XML, and While, our approach consistently achieves F1 score of 1 on different datasets. In contrast, \sotat{} and \sotaa{} 
exhibits significant f1 score variations across 
different datasets of 
\emph{json, lisp, turtle, while, xml and tinyc}. This indicates that \tool{} achieves consistent performance across different datasets corresponding to the same grammar. After carefully comparing the design of \tool{} with the other two approaches, we attribute the greater stability of \tool{}'s F1 score on most datasets to the following factor: We infer grammar from decomposed data rather than the original data. Starting with an empty grammar, we incrementally construct the grammar based on the decomposed data. Since the smaller segments in the decomposed data are highly similar and contain minimal grammar, the resulting grammar remains relatively stable.


\begin{tcolorbox}[width=3.4in]
\textbf{Answer to RQ1:}
\tool{} outperforms state-of-the-art grammar inference tools, \sotat{} and \sotaa{}, by achieving higher precision, recall, and F1 scores on most datasets. Experimental evidence also underscores the consistency of \tool{} in grammar inference. 
\end{tcolorbox}

\subsection{RQ2: Resources}
While effectiveness measures the quality of the inferred grammar, efficiency also serves as a vital factor as it assesses the operation in a timely and resource-effective manner. Thus, in this research question, we further examine the efficiency of \tool{} by reporting the time cost and computational resources. Specifically, for the sake of easy comparison, we report the following metrics in alignment with the SOTA work:
\begin{table*}
\footnotesize
\caption{Average \sotaa{} \& \sotat{} results over 10 runs on R1\& R5(c-500 is R5 of tinyc); $t$ = runtime; $t_{\text{O}}$ = oracle time; $q$ = queries sent to oracle; $m$ = peak memory usage; $\pm$ = standard deviation; \textbf{bold} = \tool{} better than \sotat{} \& \sotaa{}}\label{tab:computational_resources}
\begin{tabular}{c|cccc|cccc|cccc}
  \hline
  \multicolumn{1}{c|}{}& \multicolumn{4}{c|}{\sotaa{}}  & \multicolumn{4}{c|}{\sotat{}} &\multicolumn{4}{c}{\tool{}} 
   \\
    & $t[ks]$ & $t_{O}[ks]$  & $q[k]$ & $m[GB]$ & $t[ks]$ & $t_{O}[ks]$  & $q[k]$ & $m[GB]$& $t[ks]$ & $t_{O}[ks]$  & $q[k]$ & $m[GB]$ \\\hline
    arith&0±.0&0±.0&.5±.0&.02±.0&0±.0&.00±.0&0.40&0.02&0±.0&0±.0&.6±.0&.08±.0\\
fol&.33±.0&.2±.0&28.3±3.3&.03±.0&.15±.0&.12±.0&16.30&0.03&.67±.2&.52±.1&58.0±5.5&.08±.0\\
math&.06±.0&.04±.0&6.9±.6&.02±.0&.05±.0&.04±.0&6.30&0.02&.64±.5&.17±.1&20.9±7.4&.11±.1\\\hline
json&.17±.0&.11±.0&13.9±2.6&.05±.0&.05±.0&.04±.0&6.30&0.02&.08±.1&.05±.0&7.1±.5&.13±.0\\
lisp&10.4±2.2&2.1±1.7&40.7±25.6&18.3±22.3&1.6±.1&.96±.0&52.60&0.06&\textbf{.27±.0}&\textbf{.25±.0}&\textbf{16±.1}&.12±.1\\
turtle&.83±.3&.25±.1&29.1±8.4&.06±.0&.17±.0&.11±.0&16&0.04&\textbf{.15±.0}&.13±.0&19±.1&\textbf{.03±.0}\\
while&11.7±3.5&2.8±1.0&49.6±14.6&.88±.1&3.3±.1&.39±.0&31.60&0.13&\textbf{1.62±.1}&1.5±.1&123.8±6.5&.28±.2\\
xml&.38±.1&.15±.0&19.3±2.1&.05±.0&.22±.0&.09±.0&10.80&0.08&.94±.1&.82±.1&88±8.6&.10±.0\\\hline
curl&.44±.0&.4±.0&22.3±1.3&.05±.0&.38±.0&.35±.0&21.10&0.04&\textbf{.31±.0}&\textbf{.29±.0}&\textbf{17.5±.2}&.04±.0\\
tinyc&12.2±3.9&5.1±2.2&86.5±25.7&1.2±1.0&5.5±.2&1.9±.1&114.30&0.23&\textbf{.16±.0}&\textbf{.09±.0}&\textbf{21.8±1.1}&\textbf{.04±.0}\\
minic&15.3±5.2&13±5.1&42.6±17.9&1.4±.8&17.1±.0&15.9±.0&66.70&0.12&\textbf{8.9±1.3}&\textbf{8.7±1.2}&\textbf{32.7±5.2}&.13±.1\\\hline
c-500&37.0±13.0&15.7±7.5&137.4±48.8&3.0±3.6&11.7±1.0&3.1±.2&145.90&0.42&\textbf{.25±.1}&\textbf{.13±.0}&\textbf{23.4±2.3}&\textbf{.05±.0}\\
\hline
\end{tabular}
\end{table*}
\textbf{Runtime}: The time cost in executing the tool. 
\textbf{Oracle Time}: \sotaa{}, \sotat{} and \tool{} reply on external parser for oracle validation. Thus we measure the time cost during the oracle checking process.
\textbf{Oracle Calls}: We also calculate the number of oracle calls.
\textbf{Peak Memory}: We use the Linux `time' command to measure peak memory usage during grammar inference.

Table~\ref{tab:computational_resources} shows the performance of resource utilization for \sotaa{}, \sotat{} and \tool{} on R1\&R5 (c-500 is R5 dataset of tinyc). Overall, \tool{} takes 1.2 kiloseconds on average to process an program to do grammar inference, which is faster than that of the \sotat{} (\textit{i.e.}, on average, 3.4 kiloseconds to process a program) and the \sotaa{} (\textit{i.e.}, on average, 7.4 kiloseconds to process a program). As experimentally demonstrated by Arefin~\cite{arefin2024fast}, increasing the grammar quality will likely reduce efficiency as it requires trade-offs when handling the frequency of the validations. It explains why \tool{} takes more time on \emph{fol, math, json and xml} dataset as \tool{} takes extensive validation steps during the grammar inference process to ensure correctness. These validation steps require the use of an oracle to verify the precision of the inferred grammar rules, resulting in a slight increase of time cost. However, we argue that it is inevitable to make such trade-offs between effectiveness and efficiency. In addition, the fact that the time difference between \tool{} and \sotat{} is relatively small (\textbf{i.e.}, \emph{fol, math, json and xml}), their runtime difference is only a few hundred seconds, which is completely acceptable. On more complex datasets, our method can still maintain a fast runtime due to the data decomposition step. it suggests that \tool{} is applicable for use, despite its slightly longer execution time for some grammars, given its superior effectiveness and lower time overhead.
 
Moreover, it's worth mentioning that \tool{} spends considerably less time than \sotaa{} and \sotat{} on \emph{lisp , while , tinyc, minic and c-500} datasets. Upon scrutinizing the time expenditure of each sub-process, we ascertain that this is primarily due to \tool{}'s innovative data decomposition process, which deliberately circumvents redundant learning of identical grammars present in the original data. In essence, once \tool{} establishes a grammar rule, there is no need for further re-learning in subsequent processes, significantly enhancing efficiency.

For memory consumption, we observed a significant reduction in memory usage for \emph{tinyc} and \emph{c-500}, while for other grammars, the difference is small (at most a twofold difference). The reduced memory usage for tinyc and \emph{c-500} is primarily due to the data decomposition, which greatly reduces the runtime. Since \tool{}'s memory usage mainly stems from caching Oracle results, the reduction in runtime naturally leads to lower memory 
consumption.

\begin{tcolorbox}[width=3.4in]
\textbf{Answer to RQ2:}
\tool{} has an average runtime of 1.2 kiloseconds and memory usage of 0.1 GB for each dataset. Compared to \sotaa{} and \sotat{}, there is no significant increase in memory usage, while the average runtime significantly decreases.
\end{tcolorbox}

\subsection{RQ3:Readability}
\sotaa{}, \sotat{} do not try to simplify their inferred grammars; they merely export the grammar  when they cannot identify any further generalization steps. But \tool{} will simplify the inferred grammar after merge bubbles to speed up subsequent processing. Given a use-case involving human consumption like program understanding, it is worthwhile to investigate the readability of the generated grammar. 
Another relevant use-case is that 
overly complex grammars often contain numerous ambiguities, which can lead to difficulties for 
semantic exploration due to the presence of ambiguity.
To this end, in this research question, we leverage the following metrics to evaluate readability:

\begin{itemize} [leftmargin=*]
    \item \textbf{Grammar Size}: We calculate the count of unique non-terminals and terminals, along with the number of 
    productions and the length of each 
    production (\textit{i.e.}, the length of the right-hand-side sequence of terminals and non-terminals). The size of the grammar is then determined by summing the lengths of all its 
    productions.
    \item \textbf{Parse Time \& Memory}: We assess the total time and peak memory usage needed to parse the 1k "golden" test programs using a parser generated from a grammar inferred by \sotaa{}, \sotat{} or \tool{}.
\end{itemize}

\begin{table*}
\footnotesize
\caption{Grammar size and parse performance on R1 \& R5(c-500 is R5 of tinyc): Averages for grammars inferred in 10 runs; NT/T = unique (non-) terminals; A = rules (alternatives); l(A) = avg. rule length; S = sum of rule lengths; tP = time to parse 1k “golden” test programs; mP = peak memory while parsing; bold = \tool{} better than \sotaa{}\&\sotat{}.}\label{tab:Readbility}
\begin{tabular}{c|ccccccc|ccccccc|ccccccc}
  \hline
  \multicolumn{1}{c|}{}& \multicolumn{7}{c|}{\sotaa{}}& \multicolumn{7}{c|}{\sotat{}}  & \multicolumn{7}{c}{\tool{}} 
   \\     &  NT & A &	l(A)	&S	& T &	$t_P[ks]$ &	$m_P$ &  NT & A &	l(A)	&S	& T &	$t_P[ks]$ &	$m_P$ & NT & A &	l(A)	&S	& T&	$t_P[ks]$ &	$m_P$  \\\hline
    arith&6&31&1.2&38&16&.05&.07&6&30&1.2&36&16&.05&.07&6&\textbf{29}&1.2&\textbf{34}&16&.11&.08\\
fol&18&98&1.5&145&80&.04&.04&20&108&1.5&159&80&.04&.03&\textbf{11}&145&1.3&189&80&\textbf{.02}&.03\\
math&22&116&1.3&150&71&.07&.06&19&119&1.4&166&71&.02&.03&\textbf{18}&181&1.1&205&72&.27&.10\\\hline
json&31&158&1.4&217&74&.02&.02&17&178&1.2&208&82&.01&.02&\textbf{13}&\textbf{114}&1.2&\textbf{137}&74&.01&.02\\
lisp&22&104&9.6&738&40&.27&.05&18&95&1.7&157&40&.14&.04&\textbf{5}&\textbf{69}&1.2&\textbf{80}&40&\textbf{.05}&\textbf{.03}\\
turtle&28&122&1.6&192&67&.04&.03&15&95&1.3&124&67&.03&.03&\textbf{11}&\textbf{86}&1.2&\textbf{104}&67&\textbf{.01}&\textbf{.02}\\
while&36&91&2.1&188&18&.26&.20&24&55&2&113&18&.11&.09&\textbf{7}&\textbf{17}&3.3&\textbf{55}&18&\textbf{.03}&\textbf{.04}\\
xml&19&92&1.5&141&58&.02&.02&9&78&2&156&58&.02&.02&\textbf{7}&120&1.5&178&58&\textbf{.01}&.02\\\hline
curl&23&181&1.4&253&95&.87&.07&18&135&1.4&186&82&.09&.03&20&\textbf{121}&1.2&\textbf{140}&80&\textbf{.05}&.03\\
tinyc&56&221&2.2&460&52&.50&.07&30&150&1.8&264&51&.39&.04&\textbf{10}&\textbf{57}&1.6&\textbf{89}&49&\textbf{.03}&\textbf{.02}\\
minic&37&214&2.4&507&82&.20&.06&39&209&1.8&376&82&.15&.04&\textbf{19}&\textbf{74}&1.7&\textbf{124}&55&\textbf{.01}&\textbf{.02}\\\hline
c-500&59&240&2.4&521&52&.77&.11&41&188&1.8&345&51&.24&.05&\textbf{9}&\textbf{57}&1.6&\textbf{90}&49&\textbf{.05}&\textbf{.02}\\
\hline

\end{tabular}
\end{table*}

The results are presented in Table~\ref{tab:Readbility}, indicating among the most datasets (highlighted in bold), our approach achieves significantly higher readability score compared to previous methods. Specifically, 
not only achieving higher precision, recall and F1 score, but also \tool{}'s scores on NT, A and S are smaller than that of the other two tools. For example, regarding rule lengths, \tool{}'s grammars are smaller for 9 out of 12 languages. The most substantial difference is observed in the \emph{tinyc, minic, c-500, lisp, while} languages, where \tool{}'s grammar is less than half the size of the \sotaa{}/\sotat{} grammar. The exceptions are \emph{fol, math, and xml}, where for the math and xml languages, \tool{}'s F1 score is higher despite a slight difference in grammar size (\textit{e.g.}, xml's 178 vs. 156). In terms of \emph{fol}, although \tool{}'s grammar is larger, we argue that the overhead is relatively small (189 vs. 159), which is acceptable.

In addition, in terms of time, the fact that \tool{} spends much less time on most language datasets demonstrates that the larger grammars of \sotaa{} and \sotat{} do not improve 
parsing performance. On the contrary, for 9 out of 12 experiments, \tool{}'s parse time is less than half of \sotat{}'s and \sotaa{}'s. However, there are several outliers (\textit{i.e.}, arith, math, and json), where \tool{} has a higher F1 score despite a slight difference in time consumption. Consider memory consumption, In half of the datasets (\textit{i.e.},c-500, minic, tinyc, while, turtle and lisp), our memory usage is lower than \sotat{} and \sotaa{}, while in the other half it is higher. Overall, their memory usage does not differ significantly.

We conducted an ablation study to further evaluate the impact of each component of \tool{} on the readability of the inferred grammar. Since some components form the foundation of \tool{} and cannot function independently (\textit{e.g.}, \tool{} would not operate if either the Data Decomposition or Incremental Inference modules were removed), we opted to integrate each component into \sotat{} rather than removing them from \tool{}. This approach allowed us to assess how these additions could enhance the readability of \sotat{}’s outputs. Consequently, the experiments were designed around the following three scenarios: \sotat{} with the Data Decomposition module, \sotat{} with the Grammar Simplification module, and \sotat{} with both the Data Decomposition and Grammar Simplification modules. It is important to note that Incremental Inference is the core framework for grammar inference and not a modular component; therefore, it cannot be added to \sotat{} for experimentation. The original \sotat{} was used as the baseline for comparison.

As shown in Table \ref{tab:Ablation1}, both Data Decomposition and Grammar Simplification contribute to improved readability, as indicated by the reductions in NT, A, and S. Specifically, Grammar Simplification reduces NT by 12\%, A by 16\%, and S by 17\%, while Data Decomposition achieves reductions of 33\%, 25\%, and 28\% for these metrics, respectively. This indicates that Data Decomposition has a more significant impact on enhancing readability. When combined, Data 
Decomposition and Grammar Simplification further reduce NT, A, and S by 38\%, 29\%, and 34\%, respectively, demonstrating their effectiveness in increasing the readability of the inferred grammar.

\begin{table}
\scriptsize
\caption{Grammar Readability on R0: Averages for grammars inferred in 10 runs; \sotat{} + D = \sotat{} with Data Decomposition; \sotat{} + S = \sotat{} with Grammar Simplification; \sotat{} + D\&S = \sotat{} with Data Decomposition and Grammar Simplification. NT = unique non-terminals; A = rules (alternatives); S = sum of rule lengths.} \label{tab:Ablation1}
\begin{tabular}{c|ccc|ccc|ccc|ccc}
  \hline
  &\multicolumn{3}{c|}{\sotat{}}&\multicolumn{3}{c}{\sotat{} + D}&\multicolumn{3}{c}{\sotat{} + S}&\multicolumn{3}{c}{\sotat{} + D\&S}\\
 &  NT & A &S &  NT & A &S &  NT & A &S &  NT & A &S \\\hline
arith&6&30&36&6&28&34&5&29&35&5&27&33\\
fol&20&108&159&18&100&148&18&101&146&17&98&145\\
math&19&119&166&15&82&115&14&95&121&13&72&95\\
json&15&178&210&11&108&137&10&104&132&10&103&129\\
lisp&10&54&73&4&33&44&8&50&66&4&33&44\\
turtle&11&84&106&8&75&89&11&80&97&7&74&88\\
while&8&18&49&8&18&49&8&18&49&8&18&49\\
xml&10&82&125&10&81&125&9&73&108&9&73&103\\
curl&19&141&196&7&82&99&19&136&186&7&81&96\\
tinyc&35&164&293&18&105&177&31&130&218&18&87&133\\
minic&39&209&376&23&183&264&36&185&326&21&181&262\\

\end{tabular}
\end{table}

\begin{tcolorbox}[width=3.4in]
\textbf{Answer to RQ3:}
\tool{} not only achieves higher precision, recall, and F1 score, but also outperforms \sotat{} and \sotaa{} in readability for the majority of languages.
\end{tcolorbox}

\subsection{RQ4 : Sampling Method}

Apart from the effectiveness, efficiency, and readability of \tool{}, it is also worth investigating the sampling algorithms as they can have a huge effect on precision calculation. Actually, in our experiments, we discovered that the previous sampling methods (\textit{i.e.}, those provided by \sotaa{} and \sotat{}) could not detect deeper grammatical errors because they limit the maximum recursion depth to avoid generating excessively long samples. 
Specifically, The \sotaa{} sampling algorithm rejects samples longer than 300. When sampling grammars, if the recursion depth exceeds 5, it tends to use production rules that only contain terminal symbols. The \sotat{} sampling algorithm is similar to the \sotaa{} sampling algorithm but does not limit sample length. 

\begin{figure}
    \centering
    \includegraphics[width=1\linewidth]{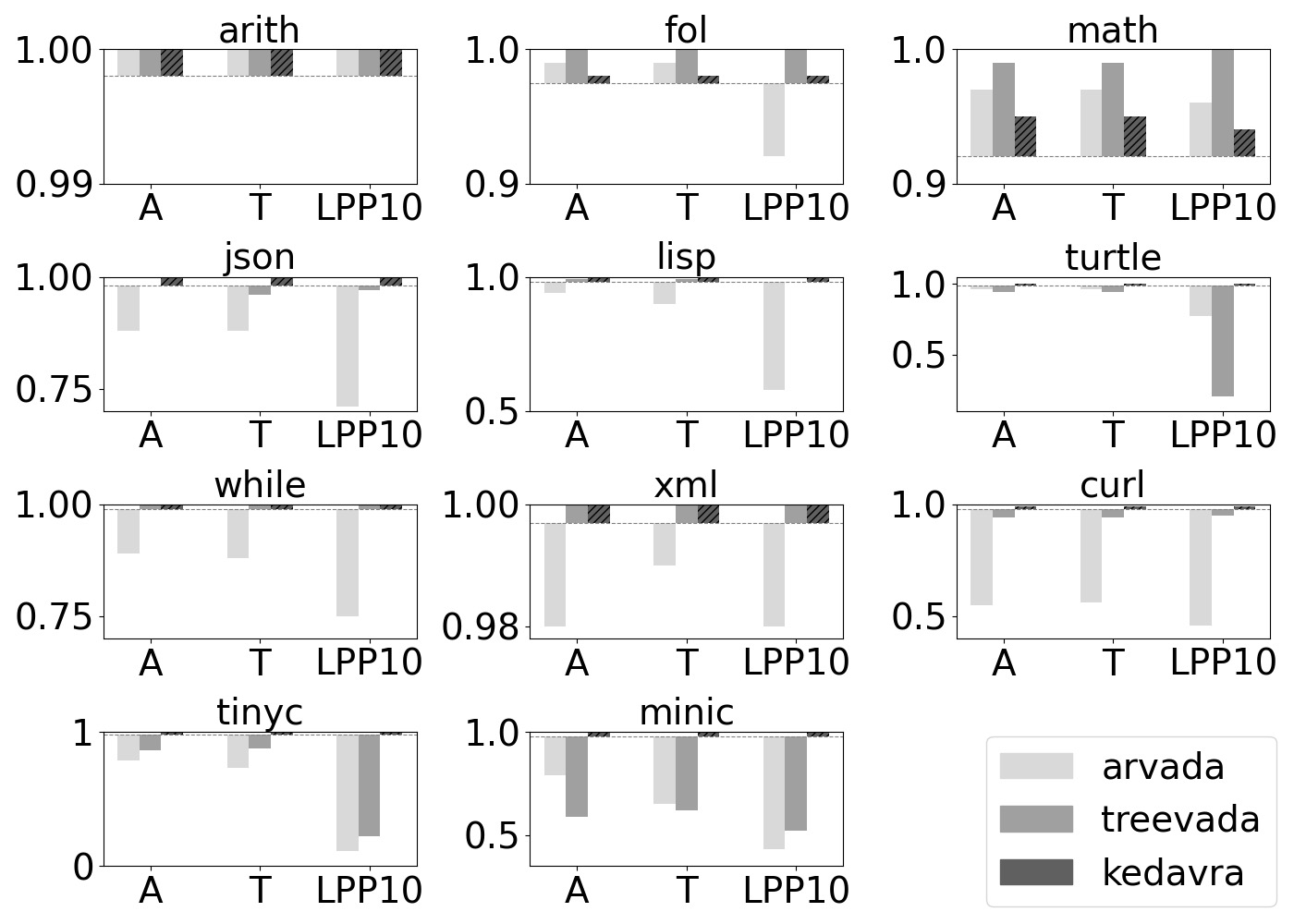}
    \caption{Precision of \sotaa{},\sotat{} and \tool{} runs on different sample methods(A = \sotaa{}'s sample algorithm,T = \sotat{}'s sample algorithm,LPP10 = LimitPerProd10). Note that the horizontal bars in each of the sub-figures are manually added as a reference to better visualize the fluctuations in the precision values of the inference algorithm across different sampling algorithms.}
    \label{fig:sample-methods}
\end{figure}

To avoid such shortcomings, We designed a more reliable algorithm that limits excessively long samples in a different way, named LimitPerProd10 (LPP10), as follows: 

\textbf{LimitPerProd10}: We limit the usage of each production rule to no more than 10 times. If a production rule is used more than 10 times, we replace the current non-terminal symbol with a precomputed string corresponding to that non-terminal symbol (to limit the maximum length of the sample). 

We then applies the three sampling algorithm on
a dataset for each grammar (R0 for arith,fol,math,minic. R1 for others). The results are presented in Figure~\ref{fig:sample-methods}, demonstrating the precision of \sotaa{}, \sotat{}, and \tool{} using three different sampling algorithms. 
Note that the horizontal bars in each of the sub-figures are manually added as a reference to better visualize the fluctuations in the precision values of the inference algorithm across different sampling algorithms.
In all the grammar datasets, \tool{} has the least variation (\textit{i.e.}, \tool{} has the shortest distance from the horizontal bar), whereas the precision of all three tools is heavily influenced by the sampling algorithm. Here, taking tinyc as an example, \sotaa{}'s minimum and maximum precision values of different sampling algorithm are respectively 0.11 and 0.79. \sotat{}'s minimum and maximum precision values of different sampling algorithm are respectively 0.22 and 0.88. However, \tool{}'s precision consistently high with minimal variation, demonstrating excellent stability.

To further explain why LPP10 outperforms the \sotaa{} and \sotat{} sampling algorithms, we manually checked some cases and discovered that \sotaa{} and \sotat{} have high precision rates with their own sampling algorithms but perform much worse with the LPP10 algorithm. This is because both the \sotaa{} and \sotat{} sampling algorithms overlook deep grammars, which are, however, exposed by the LPP10 algorithm. Therefore, the results from the LPP10 sampling algorithm better reflect reality.

Here, we proposed an example to show the differences: 
\begin{Verbatim}[frame=single,commandchars=\\\{\},codes={\catcode`$=3\catcode`_=8}]
$n_{0}:n_{1}$
$n_{1}:n_{2}$
$n_{2}:n_{3}$
$n_{3}:n_{4}$
$n_{4}:n_{5}$
$n_{5}:n_{6} | ``1" $
$n_{6}:``2" $
\end{Verbatim}
In the above example, assuming the oracle accepts only ``1'' and not ``2''. When sampling grammar, reaching $n_5$ results in maximum recursion depth for \sotaa{} and \sotat{} algorithms. In this case, these algorithms tend to choose production rules with terminal symbols, picking ``1'' instead of $n_6$, yielding a precision of 1. However, under the LPP10 sampling algorithm, each production rule's usage count is limited; for example, once $n_5 \rightarrow n_6$ is used more than 10 times, it won't be selected again. This allows us to still have the probability of selecting $n_6$ when reaching $n_5$, enabling us to detect deeper errors.

In summary, we proposed a more reliable sampling algorithm that can be used for evaluation by other grammar inference tools.
\begin{tcolorbox}[width=3.4in]
\textbf{Answer to RQ4:}
We have identified an issue with the sampling algorithms of \sotat{} and \sotaa{}, which fail to detect deep errors in complex grammars. To address this, we propose a new sampling algorithm that can more accurately evaluate grammar accuracy.
\end{tcolorbox}

\section{Discussion and Future Works}
\label{sec:discussion}

\textbf{The availability of example strings.} Current black-box context-free grammar inference approaches, including \textsc{Glade}~\cite{bastani2017synthesizing}, \sotaa{}~\cite{kulkarni2021learning}, \sotat{}~\cite{arefin2024fast}, and our \tool{}, all require both an oracle (\textit{i.e.}, a black-box parser) and the availability of example strings. 
These example strings must conform to the grammar of the target language, and the quality of the inferred grammar is significantly influenced by the quality of these strings—for example, whether the strings adequately cover most of the language’s grammar. 
However, in real-world settings, example strings are not always readily available, which restricts the practical application of existing black-box grammar inference methods. A notable attempt to circumvent the need for example strings was made by \textsc{REINAM}~\cite{wu2019reinam}, which employed an industrial symbolic execution engine to generate initial example strings before conducting grammar inference. This approach, however, is feasible only in white-box scenarios, where the internal details of the oracle are accessible, and is unsuitable for black-box grammar inference.
Looking ahead, bridging this gap constitutes a critical area for future research. Developing methodologies capable of initiating grammar inference without the need for pre-existing example strings would greatly enhance the adaptability and application of black-box grammar inference across various domains. Such advancements could significantly broaden the scope and utility of grammar inference techniques in environments constrained by the lack of data accessibility.

\textbf{The impact of the quality of the example strings.} The quality of the example strings - for example, whether the strings adequately cover most of the language’s grammar - inherently impacts the quality of the inferred grammar. \tool{} addresses this challenge through a data decomposition step, which significantly reduces the impact of example quality on the grammar generated, compared with \sotaa{} and \sotat{}. In \tool{}, the decomposed strings derived from different example strings tend to be very similar. This effect is illustrated in Figure \ref{fig:stability}, where the F1 score variation of the grammar inferred by \tool{} across different datasets is much smaller than that observed with \sotaa{} and \sotat{}.

\textbf{Performance on \emph{curl} and conventional programming languages.} \tool{} utilizes a general tokenizer that incorporates lexical rules effective across many programming languages, recognizing substrings matching the regular expression \emph{``[a-zA-Z\_][0-9a-zA-Z\_]*''} as identifiers and \emph{``[0-9]+''} as numeric tokens in most programming languages like Python, Go, Java, C/C++, \textit{etc.}. This design choice optimizes \tool{} for conventional programming languages where these patterns are prevalent.
However, these assumptions are less effective for languages like \emph{curl}, or those with identifiers that include characters beyond alphanumeric and underscores, such as \emph{Perl}, \emph{Ruby}, \emph{JavaScript}, and \emph{Scala}. \tool{}'s poor results with \emph{curl} arise from incorrect handling of URLs. For example, the tokenization step inappropriately splits the URL \emph{“http://1.g.i10//?”} into \emph{“http, :, /, /, 1, ., g, ., i10, /, /, ?”} with tokens as \emph{“t0: / / t1 t2 t2 t2 t2 / / t3”}, whereas each character after \emph{“http://”} should be treated uniformly (\textit{i.e.}, \emph{“t0 : / / t1 t1 t1 t1 t1 t1 t1 t1 t1 t1”}). This demonstrates that while \tool{} is effective with conventional programming languages whose grammars follow typical identifier rules, it struggles with non-standard token systems due to its foundational assumptions.
Addressing these specific limitations and exploring potential optimizations will be part of our future work.

\textbf{Scalability.} The scalability of \tool{} is mainly determined by the tokenization processing. Tokenization time increases quadratically with the size of the sample set due to two key steps: substring splitting, which takes $O(n)$, and merging mergeable tokens, which takes $O(n^2)$. In contrast, incremental inference, which uses decomposed strings for grammar inference, is less impacted because different samples may produce the same decomposed strings. We would like to highlight that \tool{} is less influenced by the sample set size compared to \sotaa{} and \sotat{}. For instance, in our experiments with the \emph{tinyc} datasets (Table \ref{tab:computational_resources}), specifically R1 and R5 — where R5 has a larger sample set — the runtime difference for \sotaa{}/\sotat{} is 2-3 times greater between the datasets. In contrast, the runtime increase for \tool{} from R1 to R5 is only 1.5 times, demonstrating its relative scalability under increased input sizes compared to the state of the art.


\section{Related Works}
\textbf{Black-box Grammar Inference.}
\textsc{Glade}~\cite{bastani2017synthesizing} initiated black-box grammar inference, but its performance declines with highly recursive grammars. \sotaa{}~\cite{kulkarni2021learning} advanced this by better handling recursive grammars, while \sotat{}~\cite{arefin2024fast} improved the process by optimizing for bracket structures and recursive grammar application, allowing the inference of deterministic grammars. \textsc{REINAM}~\cite{wu2019reinam}, differing from CFG, infers a target program’s PCFG\cite{jelinek1992}, first deriving an initial CFG with \textsc{Glade}, then enhancing it through reinforcement learning to adjust probabilities within the PCFG.

\textbf{Deep Learning.} 
Research\cite{sennhauser2018evaluating,bernardy2018can,yu2019learning} has assessed the limitations of RNN, in learning CFG. Despite deep learning’s inability to directly derive CFG, Yellin~\cite{yellin2021synthesizing} developed a method to extract CFG from RNN. However, results reveal substantial shortcomings of LSTM compared to \sotaa{}, likely due to the absence of \textit{active learning} \cite{angluin1981} which is utilized in \sotaa{}, \sotat{}, and \tool{}.

\textbf{Grey-box Grammar Inference.}
Grey-box inference leverages partial information from an oracle without full access to its source code. \textsc{GRIMOIRE}~\cite{blazytko2019grimoire} demonstrates this by deriving grammar-like information from oracle’s coverage data during fuzz testing, albeit not generating a direct CFG.

\textbf{White-box Grammar Inference.} 
This approach, exemplified by Lin~\cite{lin2008deriving} and \textsc{AUTOGRAM}~\cite{hoeschele2016mining}, relies on full access to oracle’s source code, using dynamic analysis and dynamic taint analysis respectively to trace inputs and deduce CFGs. Its application limited to scenarios where source code access is permissible.

\textbf{Semantics Exploration.} 
Semantic exploration has been researched by many people\cite{kovacevic2020,zhou2023semantics,steinhofel2022input}, such as tools like ISLa~\cite{steinhofel2022input}, focuses on inferring semantics within grammar by defining input constraints, either manually or mined from samples. This exploration is dependent on existing grammar, suggesting that an integrated approach with grammar derivation could significantly enhance the generation of high-quality inputs for black-box programs.

\section{Conclusions}

Inferring context-free grammars from black-box environments poses substantial challenges due to the restricted availability of example programs. Current leading methods, \sotaa{} and \sotat{}, utilize heuristic strategies to derive grammar rules from basic parse structures while examining a range of generalization sequences. These techniques, however, tend to produce grammars of lower quality and readability, a consequence of processing complete example strings which increases complexity and computational time. In response, we developed a novel methodology, \tool{}, that decomposes example strings into smaller segments for incremental grammar inference. Our method consistently outperforms existing approaches in terms of grammar precision, recall, computational efficiency, and readability, as verified by our empirical studies.

\section{Acknowledgment}

This work was supported by the Key Laboratory of Computing Power Network and Information Security, Ministry of Education under Grant No.2023ZD034.

\bibliographystyle{ACM-Reference-Format}
\bibliography{sample-base}


\begin{thebibliography}{38}


\ifx \showCODEN    \undefined \def \showCODEN     #1{\unskip}     \fi
\ifx \showDOI      \undefined \def \showDOI       #1{#1}\fi
\ifx \showISBNx    \undefined \def \showISBNx     #1{\unskip}     \fi
\ifx \showISBNxiii \undefined \def \showISBNxiii  #1{\unskip}     \fi
\ifx \showISSN     \undefined \def \showISSN      #1{\unskip}     \fi
\ifx \showLCCN     \undefined \def \showLCCN      #1{\unskip}     \fi
\ifx \shownote     \undefined \def \shownote      #1{#1}          \fi
\ifx \showarticletitle \undefined \def \showarticletitle #1{#1}   \fi
\ifx \showURL      \undefined \def \showURL       {\relax}        \fi
\providecommand\bibfield[2]{#2}
\providecommand\bibinfo[2]{#2}
\providecommand\natexlab[1]{#1}
\providecommand\showeprint[2][]{arXiv:#2}

\bibitem[Alsaeed and Young(2023)]%
        {alsaeed2023finding}
\bibfield{author}{\bibinfo{person}{Ziyad Alsaeed} {and} \bibinfo{person}{Michal Young}.} \bibinfo{year}{2023}\natexlab{}.
\newblock \showarticletitle{Finding Short Slow Inputs Faster with Grammar-Based Search}. In \bibinfo{booktitle}{\emph{Proceedings of the 32nd ACM SIGSOFT International Symposium on Software Testing and Analysis}}. \bibinfo{pages}{1068--1079}.
\newblock


\bibitem[Angluin(1981)]%
        {angluin1981}
\bibfield{author}{\bibinfo{person}{Dana Angluin}.} \bibinfo{year}{1981}\natexlab{}.
\newblock \showarticletitle{A note on the number of queries needed to identify regular languages}.
\newblock \bibinfo{journal}{\emph{Information and Control}} \bibinfo{volume}{51}, \bibinfo{number}{1} (\bibinfo{year}{1981}), \bibinfo{pages}{76--87}.
\newblock


\bibitem[Arefin et~al\mbox{.}(2024)]%
        {arefin2024fast}
\bibfield{author}{\bibinfo{person}{Mohammad~Rifat Arefin}, \bibinfo{person}{Suraj Shetiya}, \bibinfo{person}{Zili Wang}, {and} \bibinfo{person}{Christoph Csallner}.} \bibinfo{year}{2024}\natexlab{}.
\newblock \showarticletitle{Fast Deterministic Black-box Context-free Grammar Inference}. In \bibinfo{booktitle}{\emph{Proceedings of the IEEE/ACM 46th International Conference on Software Engineering}}. \bibinfo{pages}{1--12}.
\newblock


\bibitem[Aschermann et~al\mbox{.}(2019)]%
        {aschermann2019nautilus}
\bibfield{author}{\bibinfo{person}{Cornelius Aschermann}, \bibinfo{person}{Tommaso Frassetto}, \bibinfo{person}{Thorsten Holz}, \bibinfo{person}{Patrick Jauernig}, \bibinfo{person}{Ahmad-Reza Sadeghi}, {and} \bibinfo{person}{Daniel Teuchert}.} \bibinfo{year}{2019}\natexlab{}.
\newblock \showarticletitle{NAUTILUS: Fishing for Deep Bugs with Grammars.}. In \bibinfo{booktitle}{\emph{NDSS}}.
\newblock


\bibitem[Bastani et~al\mbox{.}(2017)]%
        {bastani2017synthesizing}
\bibfield{author}{\bibinfo{person}{Osbert Bastani}, \bibinfo{person}{Rahul Sharma}, \bibinfo{person}{Alex Aiken}, {and} \bibinfo{person}{Percy Liang}.} \bibinfo{year}{2017}\natexlab{}.
\newblock \showarticletitle{Synthesizing program input grammars}. In \bibinfo{booktitle}{\emph{Proceedings of the 38th ACM SIGPLAN Conference on Programming Language Design and Implementation}}. \bibinfo{publisher}{ACM}, \bibinfo{pages}{95--110}.
\newblock


\bibitem[Bernardy(2018)]%
        {bernardy2018can}
\bibfield{author}{\bibinfo{person}{Jean-Philippe Bernardy}.} \bibinfo{year}{2018}\natexlab{}.
\newblock \showarticletitle{Can recurrent neural networks learn nested recursion?}
\newblock \bibinfo{journal}{\emph{Linguistic Issues in Language Technology}}  \bibinfo{volume}{16} (\bibinfo{year}{2018}).
\newblock


\bibitem[Blazytko et~al\mbox{.}(2019)]%
        {blazytko2019grimoire}
\bibfield{author}{\bibinfo{person}{Timo Blazytko}, \bibinfo{person}{Matthew Bishop}, \bibinfo{person}{Cornelius Aschermann}, {et~al\mbox{.}}} \bibinfo{year}{2019}\natexlab{}.
\newblock \showarticletitle{{GRIMOIRE}: Synthesizing structure while fuzzing}. In \bibinfo{booktitle}{\emph{28th USENIX Security Symposium (USENIX Security 19)}}. \bibinfo{pages}{1985--2002}.
\newblock


\bibitem[Caballero et~al\mbox{.}(2007)]%
        {caballero2007polyglot}
\bibfield{author}{\bibinfo{person}{Juan Caballero}, \bibinfo{person}{Heng Yin}, \bibinfo{person}{Zhenkai Liang}, {and} \bibinfo{person}{Dawn Song}.} \bibinfo{year}{2007}\natexlab{}.
\newblock \showarticletitle{Polyglot: Automatic extraction of protocol message format using dynamic binary analysis}. In \bibinfo{booktitle}{\emph{Proceedings of the 14th ACM conference on Computer and communications security}}. \bibinfo{pages}{317--329}.
\newblock


\bibitem[Cremers and Ginsburg(1975)]%
        {cremers1975}
\bibfield{author}{\bibinfo{person}{A. Cremers} {and} \bibinfo{person}{S. Ginsburg}.} \bibinfo{year}{1975}\natexlab{}.
\newblock \showarticletitle{Context-free grammar forms}.
\newblock \bibinfo{journal}{\emph{J. Comput. System Sci.}} \bibinfo{volume}{11}, \bibinfo{number}{1} (\bibinfo{year}{1975}), \bibinfo{pages}{86--117}.
\newblock


\bibitem[Godefroid et~al\mbox{.}(2008)]%
        {godefroid2008grammar}
\bibfield{author}{\bibinfo{person}{Patrice Godefroid}, \bibinfo{person}{Adam Kiezun}, {and} \bibinfo{person}{Michael~Y Levin}.} \bibinfo{year}{2008}\natexlab{}.
\newblock \showarticletitle{Grammar-based whitebox fuzzing}. In \bibinfo{booktitle}{\emph{Proceedings of the 29th ACM SIGPLAN conference on programming language design and implementation}}. \bibinfo{pages}{206--215}.
\newblock


\bibitem[Gopinath and Zeller(2019)]%
        {gopinath2019building}
\bibfield{author}{\bibinfo{person}{Rahul Gopinath} {and} \bibinfo{person}{Andreas Zeller}.} \bibinfo{year}{2019}\natexlab{}.
\newblock \showarticletitle{Building fast fuzzers}.
\newblock \bibinfo{journal}{\emph{arXiv preprint arXiv:1911.07707}} (\bibinfo{year}{2019}).
\newblock


\bibitem[Gosling et~al\mbox{.}(2022)]%
        {java}
\bibfield{author}{\bibinfo{person}{James Gosling}, \bibinfo{person}{Bill Joy}, \bibinfo{person}{Guy Steele}, \bibinfo{person}{Gilad Bracha}, \bibinfo{person}{Alex Buckley}, \bibinfo{person}{Daniel Smith}, {and} \bibinfo{person}{Gavin Bierman}.} \bibinfo{year}{2022}\natexlab{}.
\newblock \bibinfo{booktitle}{\emph{The Java Language Specification: Java SE 19 Edition}}.
\newblock \bibinfo{publisher}{Oracle}.
\newblock


\bibitem[Höschele and Zeller(2016)]%
        {hoeschele2016mining}
\bibfield{author}{\bibinfo{person}{Matthias Höschele} {and} \bibinfo{person}{Andreas Zeller}.} \bibinfo{year}{2016}\natexlab{}.
\newblock \showarticletitle{Mining input grammars from dynamic taints}. In \bibinfo{booktitle}{\emph{Proc. 31st IEEE/ACM International Conference on Automated Software Engineering (ASE)}}. ACM, \bibinfo{pages}{720--725}.
\newblock


\bibitem[Jelinek et~al\mbox{.}(1992)]%
        {jelinek1992}
\bibfield{author}{\bibinfo{person}{Frederick Jelinek}, \bibinfo{person}{John~D. Lafferty}, {and} \bibinfo{person}{Robert~L. Mercer}.} \bibinfo{year}{1992}\natexlab{}.
\newblock \bibinfo{booktitle}{\emph{Basic methods of probabilistic context free grammars}}.
\newblock \bibinfo{publisher}{Springer Berlin Heidelberg}.
\newblock


\bibitem[Kim et~al\mbox{.}(2013)]%
        {kim2013specification}
\bibfield{author}{\bibinfo{person}{Tae-Woong Kim}, \bibinfo{person}{Tae-Gong Kim}, {and} \bibinfo{person}{Jai-Hyun Seu}.} \bibinfo{year}{2013}\natexlab{}.
\newblock \showarticletitle{Specification and automated detection of code smells using OCL}.
\newblock \bibinfo{journal}{\emph{International Journal of Software Engineering and Its Applications}} \bibinfo{volume}{7}, \bibinfo{number}{4} (\bibinfo{year}{2013}), \bibinfo{pages}{35--44}.
\newblock


\bibitem[Kovačević et~al\mbox{.}(2020)]%
        {kovacevic2020}
\bibfield{author}{\bibinfo{person}{Željko Kovačević}, \bibinfo{person}{Marjan Mernik}, \bibinfo{person}{Matej Ravber}, {et~al\mbox{.}}} \bibinfo{year}{2020}\natexlab{}.
\newblock \showarticletitle{From grammar inference to semantic inference—An evolutionary approach}.
\newblock \bibinfo{journal}{\emph{Mathematics}} \bibinfo{volume}{8}, \bibinfo{number}{5} (\bibinfo{year}{2020}), \bibinfo{pages}{816}.
\newblock
\urldef\tempurl%
\url{https://doi.org/10.3390/math8050816}
\showDOI{\tempurl}


\bibitem[Kulkarni et~al\mbox{.}(2021)]%
        {kulkarni2021learning}
\bibfield{author}{\bibinfo{person}{Neil Kulkarni}, \bibinfo{person}{Caroline Lemieux}, {and} \bibinfo{person}{Koushik Sen}.} \bibinfo{year}{2021}\natexlab{}.
\newblock \showarticletitle{Learning highly recursive input grammars}. In \bibinfo{booktitle}{\emph{2021 36th IEEE/ACM International Conference on Automated Software Engineering (ASE)}}. IEEE, \bibinfo{pages}{456--467}.
\newblock


\bibitem[Köppe et~al\mbox{.}(2022)]%
        {cpp}
\bibfield{author}{\bibinfo{person}{Thomas Köppe} {et~al\mbox{.}}} \bibinfo{year}{2022}\natexlab{}.
\newblock \bibinfo{booktitle}{\emph{Working Draft, Standard for Programming Language {C++}}}.
\newblock \bibinfo{type}{{T}echnical {R}eport} N4917. \bibinfo{institution}{ISO/IEC}.
\newblock


\bibitem[Lin and Zhang(2008)]%
        {lin2008deriving}
\bibfield{author}{\bibinfo{person}{Zhiqiang Lin} {and} \bibinfo{person}{Xiangyu Zhang}.} \bibinfo{year}{2008}\natexlab{}.
\newblock \showarticletitle{Deriving input syntactic structure from execution}. In \bibinfo{booktitle}{\emph{Proceedings of the 16th ACM SIGSOFT International Symposium on Foundations of Software Engineering}}. \bibinfo{pages}{83--93}.
\newblock


\bibitem[Marchal(2002)]%
        {marchal2002}
\bibfield{author}{\bibinfo{person}{Benoît Marchal}.} \bibinfo{year}{2002}\natexlab{}.
\newblock \bibinfo{booktitle}{\emph{{XML by Example}}}.
\newblock \bibinfo{publisher}{Que Publishing}.
\newblock


\bibitem[Moonen(2001)]%
        {moonen2001generating}
\bibfield{author}{\bibinfo{person}{Leon Moonen}.} \bibinfo{year}{2001}\natexlab{}.
\newblock \showarticletitle{Generating robust parsers using island grammars}. In \bibinfo{booktitle}{\emph{Proceedings eighth working conference on reverse engineering}}. IEEE, \bibinfo{pages}{13--22}.
\newblock


\bibitem[Nagy and Cleve(2017)]%
        {nagy2017static}
\bibfield{author}{\bibinfo{person}{Csaba Nagy} {and} \bibinfo{person}{Anthony Cleve}.} \bibinfo{year}{2017}\natexlab{}.
\newblock \showarticletitle{A static code smell detector for SQL queries embedded in Java code}. In \bibinfo{booktitle}{\emph{2017 IEEE 17th International Working Conference on Source Code Analysis and Manipulation (SCAM)}}. IEEE, \bibinfo{pages}{147--152}.
\newblock


\bibitem[Narayan et~al\mbox{.}(2015)]%
        {narayan2015survey}
\bibfield{author}{\bibinfo{person}{John Narayan}, \bibinfo{person}{Sandeep Shukla}, {and} \bibinfo{person}{Charles Clancy}.} \bibinfo{year}{2015}\natexlab{}.
\newblock \showarticletitle{A survey of automatic protocol reverse engineering tools}.
\newblock \bibinfo{journal}{\emph{Comput. Surveys}} \bibinfo{volume}{48}, \bibinfo{number}{3} (\bibinfo{year}{2015}), \bibinfo{pages}{1--36}.
\newblock
\urldef\tempurl%
\url{https://doi.org/10.1145/2732254}
\showDOI{\tempurl}


\bibitem[Nguyen et~al\mbox{.}(2020)]%
        {nguyen2020mofuzz}
\bibfield{author}{\bibinfo{person}{Hoang~Lam Nguyen}, \bibinfo{person}{Nebras Nassar}, \bibinfo{person}{Timo Kehrer}, {and} \bibinfo{person}{Lars Grunske}.} \bibinfo{year}{2020}\natexlab{}.
\newblock \showarticletitle{Mofuzz: A fuzzer suite for testing model-driven software engineering tools}. In \bibinfo{booktitle}{\emph{Proc. 35th IEEE/ACM International Conference on Automated Software Engineering}}. \bibinfo{pages}{1103--1115}.
\newblock


\bibitem[Oda et~al\mbox{.}(2015)]%
        {oda2015learning}
\bibfield{author}{\bibinfo{person}{Yusuke Oda}, \bibinfo{person}{Hiroyuki Fudaba}, \bibinfo{person}{Graham Neubig}, \bibinfo{person}{Hideaki Hata}, \bibinfo{person}{Sakriani Sakti}, \bibinfo{person}{Tomoki Toda}, {and} \bibinfo{person}{Satoshi Nakamura}.} \bibinfo{year}{2015}\natexlab{}.
\newblock \showarticletitle{Learning to generate pseudo-code from source code using statistical machine translation}. In \bibinfo{booktitle}{\emph{Proc. 30th IEEE/ACM International Conference on Automated Software Engineering (ASE)}}. IEEE, \bibinfo{pages}{574--584}.
\newblock


\bibitem[Sakakibara(1990)]%
        {sakakibara1990}
\bibfield{author}{\bibinfo{person}{Yasubumi Sakakibara}.} \bibinfo{year}{1990}\natexlab{}.
\newblock \showarticletitle{Learning context-free grammars from structural data in polynomial time}.
\newblock \bibinfo{journal}{\emph{Theoretical Computer Science}} \bibinfo{volume}{76}, \bibinfo{number}{2-3} (\bibinfo{year}{1990}), \bibinfo{pages}{223--242}.
\newblock


\bibitem[Sakakibara(1992)]%
        {sakakibara1992}
\bibfield{author}{\bibinfo{person}{Yasubumi Sakakibara}.} \bibinfo{year}{1992}\natexlab{}.
\newblock \showarticletitle{Efficient learning of context-free grammars from positive structural examples}.
\newblock \bibinfo{journal}{\emph{Information and Computation}} \bibinfo{volume}{97}, \bibinfo{number}{1} (\bibinfo{year}{1992}), \bibinfo{pages}{23--60}.
\newblock


\bibitem[Sakakibara and Muramatsu(2000)]%
        {sakakibara2000}
\bibfield{author}{\bibinfo{person}{Yasubumi Sakakibara} {and} \bibinfo{person}{Hiroshi Muramatsu}.} \bibinfo{year}{2000}\natexlab{}.
\newblock \showarticletitle{Learning context-free grammars from partially structured examples}. In \bibinfo{booktitle}{\emph{Grammatical Inference: Algorithms and Applications: 5th International Colloquium, ICGI 2000, Lisbon, Portugal, September 11-13, 2000. Proceedings}}. \bibinfo{publisher}{Springer Berlin Heidelberg}, \bibinfo{pages}{229--240}.
\newblock


\bibitem[Sennhauser and Berwick(2018)]%
        {sennhauser2018evaluating}
\bibfield{author}{\bibinfo{person}{Luzi Sennhauser} {and} \bibinfo{person}{Robert~C Berwick}.} \bibinfo{year}{2018}\natexlab{}.
\newblock \showarticletitle{Evaluating the ability of LSTMs to learn context-free grammars}.
\newblock \bibinfo{journal}{\emph{arXiv preprint arXiv:1811.02611}} (\bibinfo{year}{2018}).
\newblock


\bibitem[Smith(2015)]%
        {smith2015}
\bibfield{author}{\bibinfo{person}{Ben Smith}.} \bibinfo{year}{2015}\natexlab{}.
\newblock \bibinfo{booktitle}{\emph{{Beginning JSON}}}.
\newblock \bibinfo{publisher}{Apress}.
\newblock


\bibitem[Srivastava and Payer(2021)]%
        {srivastava2021gramatron}
\bibfield{author}{\bibinfo{person}{Prashast Srivastava} {and} \bibinfo{person}{Mathias Payer}.} \bibinfo{year}{2021}\natexlab{}.
\newblock \showarticletitle{Gramatron: Effective grammar-aware fuzzing}. In \bibinfo{booktitle}{\emph{Proceedings of the 30th ACM SIGSOFT International Symposium on Software Testing and Analysis}}. \bibinfo{pages}{244--256}.
\newblock


\bibitem[Steinh{\"o}fel and Zeller(2022)]%
        {steinhofel2022input}
\bibfield{author}{\bibinfo{person}{D. Steinh{\"o}fel} {and} \bibinfo{person}{A. Zeller}.} \bibinfo{year}{2022}\natexlab{}.
\newblock \showarticletitle{Input invariants}. In \bibinfo{booktitle}{\emph{Proceedings of the 30th ACM Joint European Software Engineering Conference and Symposium on the Foundations of Software Engineering}}. \bibinfo{pages}{583--594}.
\newblock


\bibitem[Wang et~al\mbox{.}(2019)]%
        {wang2019superion}
\bibfield{author}{\bibinfo{person}{Junjie Wang}, \bibinfo{person}{Bihuan Chen}, \bibinfo{person}{Lei Wei}, {and} \bibinfo{person}{Yang Liu}.} \bibinfo{year}{2019}\natexlab{}.
\newblock \showarticletitle{Superion: Grammar-aware greybox fuzzing}. In \bibinfo{booktitle}{\emph{2019 IEEE/ACM 41st International Conference on Software Engineering (ICSE)}}. IEEE, \bibinfo{pages}{724--735}.
\newblock


\bibitem[Wu et~al\mbox{.}(2019)]%
        {wu2019reinam}
\bibfield{author}{\bibinfo{person}{Z. Wu}, \bibinfo{person}{E. Johnson}, \bibinfo{person}{W. Yang}, {et~al\mbox{.}}} \bibinfo{year}{2019}\natexlab{}.
\newblock \showarticletitle{REINAM: reinforcement learning for input-grammar inference}. In \bibinfo{booktitle}{\emph{Proceedings of the 2019 27th ACM Joint Meeting on European Software Engineering Conference and Symposium on the Foundations of Software Engineering}}. \bibinfo{pages}{488--498}.
\newblock


\bibitem[Yellin and Weiss(2021)]%
        {yellin2021synthesizing}
\bibfield{author}{\bibinfo{person}{Daniel~M Yellin} {and} \bibinfo{person}{Gail Weiss}.} \bibinfo{year}{2021}\natexlab{}.
\newblock \showarticletitle{Synthesizing context-free grammars from recurrent neural networks}. In \bibinfo{booktitle}{\emph{International Conference on Tools and Algorithms for the Construction and Analysis of Systems}}. \bibinfo{publisher}{Springer}, \bibinfo{pages}{351--369}.
\newblock


\bibitem[Yu et~al\mbox{.}(2019)]%
        {yu2019learning}
\bibfield{author}{\bibinfo{person}{Xiang Yu}, \bibinfo{person}{Ngoc~Thang Vu}, {and} \bibinfo{person}{Jonas Kuhn}.} \bibinfo{year}{2019}\natexlab{}.
\newblock \showarticletitle{Learning the Dyck language with attention-based Seq2Seq models}. In \bibinfo{booktitle}{\emph{Proceedings of the 2019 ACL Workshop BlackboxNLP: Analyzing and Interpreting Neural Networks for NLP}}. \bibinfo{pages}{138--146}.
\newblock


\bibitem[yu~Guo et~al\mbox{.}(2022)]%
        {javascript}
\bibfield{author}{\bibinfo{person}{Shu yu Guo}, \bibinfo{person}{Michael Ficarra}, {and} \bibinfo{person}{Kevin Gibbons}.} \bibinfo{year}{2022}\natexlab{}.
\newblock \bibinfo{booktitle}{\emph{ECMAScript 2022 Language Specification (13th ed.)}}.
\newblock \bibinfo{type}{Technical Report} ECMA-262. \bibinfo{institution}{Ecma International}.
\newblock


\bibitem[Zhou et~al\mbox{.}(2023)]%
        {zhou2023semantics}
\bibfield{author}{\bibinfo{person}{C. Zhou}, \bibinfo{person}{Q. Zhang}, \bibinfo{person}{L. Guo}, {et~al\mbox{.}}} \bibinfo{year}{2023}\natexlab{}.
\newblock \showarticletitle{Towards better semantics exploration for browser fuzzing}. In \bibinfo{booktitle}{\emph{Proceedings of the ACM on Programming Languages}}, Vol.~\bibinfo{volume}{7}. \bibinfo{pages}{604--631}.
\newblock


\end{thebibliography}

\end{document}